\documentclass[aps,prb,superscriptaddress,floatfix,notitlepage,twocolumn,nofootinbib]{revtex4-2}
\usepackage[justification=Justified]{caption}
\usepackage{amsfonts}
\usepackage{amsmath}
\usepackage{amssymb}
\usepackage{braket}
\usepackage{hyperref}
\usepackage{graphicx,dsfont}
\usepackage{subcaption}
\usepackage[mode=buildnew]{standalone}
\usepackage{xcolor}
\hypersetup{
   colorlinks,
   linkcolor={blue!70!black},
   citecolor={blue!70!black},
   urlcolor= {blue!70!black}
}

\def\beq{\begin{equation}}
\def\eeq{\end{equation}}

\def\l{\left}
\def\r{\right}

\def\bx{\begin{pmatrix}}
\def\ex{\end{pmatrix}}

\begin{document}

\author{Will Holdhusen}
\affiliation{Department of Physics, Indiana University, Bloomington, IN 47405, USA}

\author{Daniel Huerga}
\affiliation{Stewart Blusson Quantum Matter Institute, University of British Columbia, Vancouver, BC V6T 1Z4, Canada}

\author{Gerardo Ortiz}
\affiliation{Department of Physics, Indiana University, Bloomington, IN 47405, USA}
\affiliation{Institute for Quantum Computing, University of Waterloo, Waterloo, ON N2L 3G1, Canada}

\begin{abstract}
The Kitaev spin liquid, stabilized as the ground state of the Kitaev honeycomb model, is a paradigmatic example of a topological $\mathbb{Z}_2$ quantum spin liquid.
The fate of the Kitaev spin liquid in presence of an external magnetic field is a topic of current interest due to experiments, which apparently unveil a $\mathbb{Z}_2$ topological phase in the so-called Kitaev materials, and theoretical studies predicting the emergence of an intermediate quantum phase of debated nature before the appearance of a trivial partially polarized phase. 
In this work, we employ hierarchical mean-field theory, an algebraic and numerical method based on the use of clusters preserving relevant symmetries and short-range quantum correlations, to investigate the quantum phase diagram of the antiferromagnetic Kitaev's model in a [111] field.  
By using clusters of 24 sites, we predict that the Kitaev spin liquid transits through two intermediate phases, characterized by stripe and chiral order, respectively, before entering the trivial partially polarized phase, differing from previous studies.
We assess our results by performing exact diagonalization and computing the scaling of different observables, including the many-body Chern number and other topological quantities, thus establishing hierarchical mean-field theory as a method to study topological quantum spin liquids.
\end{abstract}

\title{Emergent magnetic order in the antiferromagnetic Kitaev model with a [111] field}
\maketitle

\section{Introduction}

Quantum spin liquids (QSLs) are featureless phases of matter resulting from competing interactions among elementary magnetic degrees of freedom. 
While no consensus exists on the precise operational characterization of a QSL, commonly accepted defining properties include translational and rotational invariance, the absence of long-range (Landau) magnetic order, and incipient topological order \cite{knolle2019, savary2016}. 
Perhaps the most agreed-upon example of a QSL is found in the ground state of the Kitaev honeycomb model (KHM) \cite{kitaev2006}.
This exactly-solvable model, introduced in 2006,  provides an archetypal example of a topological $\mathbb{Z}_2$ QSL (the Kitaev spin liquid, KSL) hosting non-abelian anyons, and constitutes a potential resource for quantum information processing.
Given the seemingly unphysical interactions constituting the KHM, indications that physical realizations may be possible in the so-called Kitaev materials are surprising. While these materials, most famously $\alpha$-RuCl$_3$, exhibit antiferromagnetic ordering, an applied magnetic field suppresses the order and uncovers KSL-like fractionalization \cite{plumb2014, sears2015, sandilands2015}. Importantly, interactions beyond those in Kitaev's exactly-solvable Hamiltonian are also present in these materials \cite{chaloupka2010, kim2016, janssen2017, trebst2022}.

The success of this field has led to further interest in the fate of the KSL in an extended KHM \cite{janssen2016, gordon2019, yang2020, sorensons2021, yang2021, zhang2021a}. Numerical simulations have shown that even the simple application of a uniform magnetic field leads to unconventional behavior: with antiferromagnetic Kitaev interactions, the KSL persists up to a relatively high field strength before transitioning into an apparently featureless intermediate phase, which in turn transitions into a partially-polarized state at yet higher fields \cite{zhu2018, gohlke2018}.
Since even the presence of a uniform magnetic field takes the KHM outside of its exactly-solvable regime, numerical simulations, variational methods, and mean-field theories must be relied upon to approach the model and uncover emergent phases of matter. As these techniques each have their own biases and drawbacks, a robust understanding of the model's quantum phase diagram requires a holistic approach.

Until recently, the majority of studies have focused on results derived from exact-diagonalization (ED) \cite{zhu2018, hickey2019, ronquillo2019, pradhan2020} and density-matrix renormalization group (DMRG) calculations \cite{zhu2018, gohlke2018, patel2019, pradhan2020}. These calculations seem to consistently predict a gapless U(1) QSL in the intermediate phase. Recent work with an effective mean-field theory over Majorana fermion degrees of freedom has provided an alternative identification of the intermediate phase as a gapped topological QSL belonging to Kitaev's 16-fold way \cite{zhang2021}, agreeing with a prediction from variational Monte Carlo \cite{jiang2020}.

Here, we provide an augmenting perspective by approaching the problem with hierarchical mean-field theory (HMFT). HMFT is a mean-field theory based on cluster degrees of freedom  preserving relevant symmetries and quantum correlations of the Hamiltonian \cite{ortiz2003,ortiz2004}. 
This method provides a simulation of the thermodynamic limit and a variational upper bound to the exact ground state energy, approaching the exact result through finite-scaling analysis with increasing cluster size \cite{isaev2009}.
HMFT has proven successful in recovering the phase diagram of systems with competing long-range orders (LROs) and quantum paramagnetic phases \cite{isaev2009, isaev2009a, huerga2016}, including the prediction of a devil's staircase of valence-bond crystals in the kagome Heisenberg antiferromagnet \cite{huerga2016} confirmed in later experiments \cite{okuma2019}.

\begin{figure}[t]
    \includegraphics[width=\linewidth]{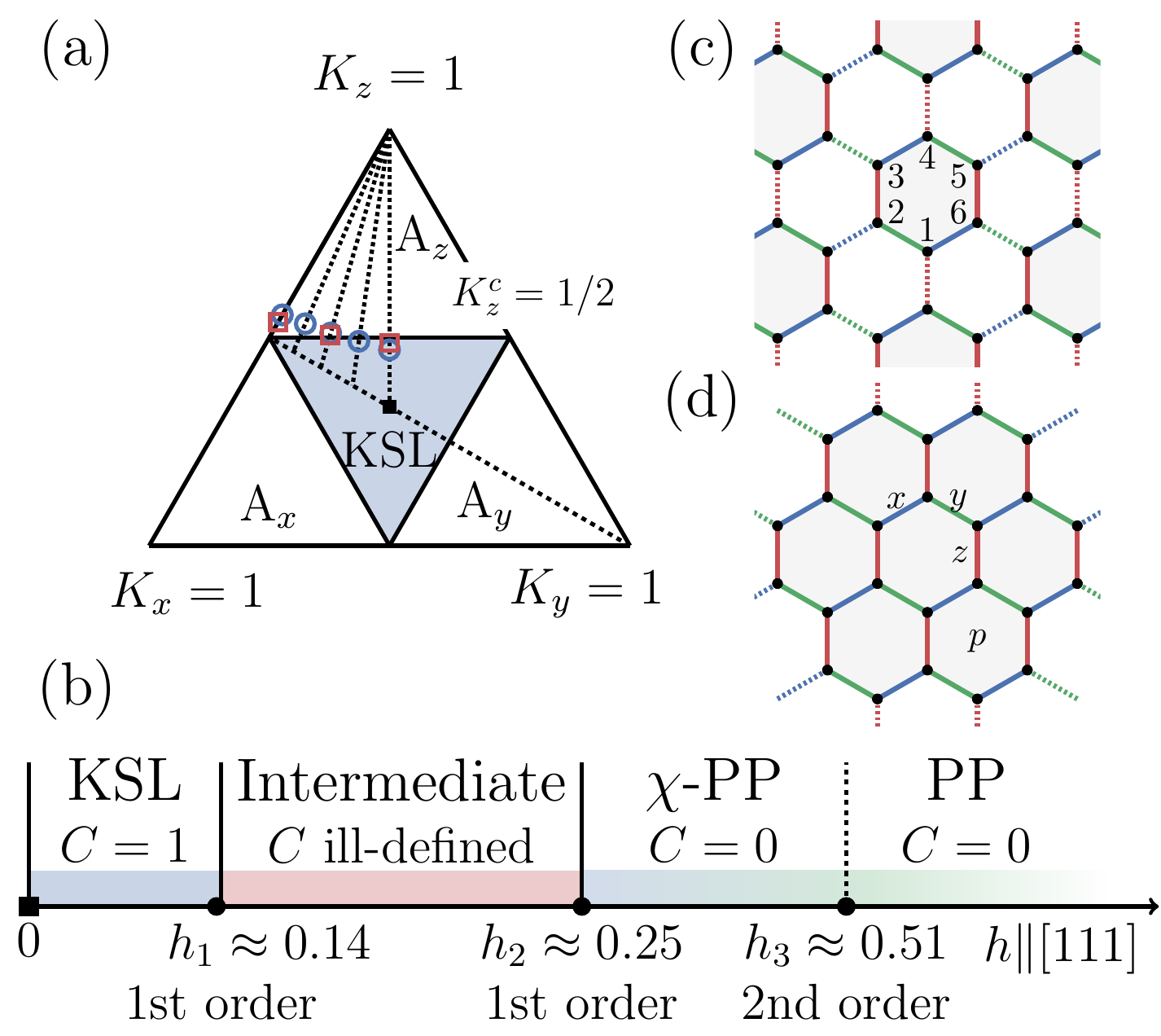}
\caption{
(a) Quantum phase diagram of the KHM as obtained with 6- and 24-site HMFT (blue circles and red squares, respectively) as compared to the exact solution (continuous black). 
(b) Schematic quantum phase diagram of the KHM at $K_x$=$K_y$=$K_z$=$1$ under external magnetic field $h$$\parallel$$[111]$ as obtained with HMFT and ED, including the Kitaev spin liquid (KSL), intermediate, and chiral and trivial partially polarized phases ($\chi$-PP and PP, respectively).
Phase boundaries correspond to 24-site HMFT results and Chern numbers $C$ have been computed with 24-site ED. 
(c) Cluster used for 6-site HMFT. Indexing corresponds to the plaquette flux $W_p$ (Eq. \eqref{eqn:wp}).
(d) Cluster used for 24-site HMFT with KHM bond directions labelled. The index $p$ indicates the center of one of the plaquettes on which $W_p$ is defined.
}
\label{fig:phase_diagrams}
\end{figure}

The present work represents the first application of HMFT to a model hosting topological order. 
We begin by simulating the exactly-solvable KHM to assess our approach, obtaining a phase diagram in good agreement with the exact solution as shown in Fig. \ref{fig:phase_diagrams}(a).
Then, we uncover the phase diagram upon application of a magnetic field along the $[111]$ direction (Fig. \ref{fig:phase_diagrams}(b)) and utilize ED to provide support to the new results we obtain from HMFT.
While HMFT confirms the presence of an intermediate phase appearing in a $[111]$ field, the phase diagram we obtain with this method has important distinctions from earlier results.

First, we find that the intermediate phase spontaneously breaks the rotational symmetry of the KHM Hamiltonian due to an otherwise suboptimal mean-field configuration crossing below the KSL mean-field energy.
In this phase, we find indications of stripe magnetic order and, therefore, of a phase with long-range order (LRO) rather than the featurelessness characteristic of a QSL, per its usual definition \cite{knolle2019, savary2016}.
Second, we find a chiral partially-polarized ($\chi$-PP) phase occurring between the intermediate and the trivial partially-polarized (PP) phases. The $\chi$-PP phase has gone unnoticed in previous studies based on ED \cite{zhu2018, hickey2019, ronquillo2019, pradhan2020} and DMRG \cite{zhu2018, gohlke2018, patel2019, pradhan2020}.
This newly-uncovered phase is characterized by a sublattice chiral order parameter and is separated from the PP phase by a second-order phase transition. 

The remainder of this  introduction outlines the organization of the manuscript.
We first review the exactly-solvable Kitaev honeycomb model (KHM) at zero field in Sec. \ref{sec:model} and the methods used (HMFT and ED) in Sec. \ref{sec:methods}.
In Sec. \ref{sec:results}, we present our results on the HMFT approach to the KHM in a $[111]$ field, making particular emphasis on our new results: the emergence of stripe order in the intermediate phase and the novel $\chi$-PP phase. In this section, we also make use of ED to assess the validity of the HMFT results.
Finally, in Sec. \ref{sec:conclusion} we conclude with remarks examining the consequences of our study and opportunities for future work building on and further testing the resulting predictions.

\section{Model}
\label{sec:model}
The $S$=1/2 KHM \cite{kitaev2006},
$H$=$\sum_\gamma \sum_{\langle i,j\rangle^\gamma}  K_\gamma S_i^\gamma S_j^\gamma$,
characterized by bond-dependent nearest-neighbor $\langle i,j\rangle^\gamma$ ($\gamma$$\in$$\{x,y,z\}$) interactions, is the paradigmatic model stabilizing QSL phases characterized by topological order.
Its exact solution is recovered upon mapping the $S$=1/2 spins to Majorana fermions coupled to a $\mathbb Z_2$ gauge field \cite{kitaev2006,nussinov2009}.
Its implications to quantum computation have made the search for Kitaev interactions in materials a consequential line of research.
Upon applying an external magnetic field $h$ along the $[111]$ direction to the KHM,
\beq
H = \sum_\gamma \sum_{\langle i,j\rangle^\gamma} K_\gamma S_i^\gamma S_j^\gamma - h\sum_i \left(S_i^x+S_i^y+S_i^z\right),
\label{eqn:KHM}
\eeq
Kitaev showed that a gap opens for $h$$\ll$1, revealing a topological non-trivial ground state characterized by Chern number $C$=$\pm 1$ in the KSL phase, making the system a resource for topological quantum computation via the braiding of its non-abelian anyon excitations \cite{kitaev2006}.

At $h$=$0$, a set of plaquette observables defined on the dual lattice,
i.e. at each 6-site hexagon $p$ of the honeycomb lattice (see Fig. 
\ref{fig:phase_diagrams}(c)),
\beq
W_p = 2^6 S_1^z S_2^x S_3^y S_4^z S_5^x S_6^y,
\label{eqn:wp}
\eeq
commutes with the Hamiltonian \eqref{eqn:KHM}, rendering it exactly-solvable.
The ground state has a well defined value $\braket{W_p}$=1 for all plaquettes in the lattice  \cite{kitaev2006}.
The quantum phase diagram of the model comprises four phases:
three topologically trivial, gapped phases (dubbed A$_\gamma$), occurring when
$
|K_\gamma| > |K_\alpha| + |K_\beta|,$ with $\{\alpha,\beta,\gamma\}\in\{x,y,z\}
$,
and a gapless (at $h$=0) topological phase (the KSL), otherwise. Figure  \ref{fig:phase_diagrams}(a) illustrates this phase diagram projected onto a surface upon which $K_x$+$K_y$+$K_z$=1.

The transition of the topological KSL phase towards a trivial partially polarized phase (PP) emerging at large magnetic fields $h$ is currently under scrutiny.
Specifically, recent numerical analysis of the antiferromagnetic ($K_\gamma > 0$) KHM has argued for the transition of the KSL phase to an intermediate, finite-field QSL phase whose fundamental nature is under debate, with arguments in favor of a gapless U(1) QSL \cite{zhu2018, gohlke2018, hickey2019, patel2019, ronquillo2019, pradhan2020} or a topological gapped QSL \cite{zhang2021, jiang2020}.

We now briefly discuss real-space symmetries of the KHM, as an analysis of these symmetries is important in order to understand when they are broken.
Due to its anisotropic bond-dependent interactions, the KHM does not preserve the C$_6$ rotational symmetry of the honeycomb lattice. 
Even for the most symmetric set of couplings ($K_x$=$K_y$=$K_z$, occurring in the KSL phase), a rotation of the lattice by $\pi/3$ about the center of an hexagon (C$_6$) must be accompanied by a $2\pi/3$ rotation of the Bloch sphere about the $[111]$ axis (C$_3^S$), resulting in a combined C$_6$$\times$C$_3^S$ symmetry.\footnote{Aligning the spin axes along real-space directions (such that a lattice rotation also rotates the spins) results in a different symmetry classification \cite{ronquillo2020}.} This remains a symmetry of the model under the application of a magnetic field, so long as it is applied in the [111] direction.

\section{Methods}
\label{sec:methods}
\subsection{Hierarchical mean-field theory}
Hierarchical mean-field theory (HMFT) is an algebraic framework and numerical method to 
approach models of strongly-correlated systems with frustrating interactions.
The main idea of the method builds upon the identification of relevant degrees of freedom (generally, clusters of the original degrees of freedom) containing the necessary quantum correlations required to unveil the phases emerging in the system under study.
By utilizing the exact mappings relating the algebras of the original and new degrees of freedom, we may encounter emerging symmetries and exact solutions \cite{batista2004} or, in their absence, utilize mean-field approaches \cite{ortiz2003, ortiz2004}. 
Under the assumption that deep within a non-critical phase, the characteristic correlation length has a finite length of few sites, we generically make use of clusters containing $N_c$ sites that uniformly tile the lattice and preserve as many symmetries of the original Hamiltonian as possible.
Thus, quantum correlations within the cluster are described from the onset, while the remaining interactions among clusters may be approximated by different mean-field approaches.

The lowest-order mean-field approximation consists of a simple product of clusters, i.e. a uniform cluster-Gutzwiller ansatz (CGA),
\beq
\ket{\Psi} = \bigotimes_\mathbf {R} \ket{\psi_\mathbf{R}},
\label{eqn:cga}
\eeq
where clusters at superlattice sites $\mathbf{R}$ are in the same state, $\ket{\psi_\mathbf{R}}$=$\sum_{\lbrace \sigma\rbrace} w_{\lbrace \sigma\rbrace}\ket{\lbrace\sigma\rbrace}$, and $w_{\lbrace \sigma\rbrace}$ are variational parameters in the basis of spin configurations of the cluster, $\lbrace \sigma\rbrace$.
These variational parameters are optimized upon minimization of the energy density in the thermodynamic limit, 
\beq
e=\frac{1}{M N_c}\frac{\bra{\Psi} H \ket{\Psi}}{\braket{\Psi|\Psi}},
\label{eqn:energy}
\eeq
where $M$ is the total number of clusters in the superlattice. From a technical standpoint, minimization of Eq. \eqref{eqn:energy} is equivalent to performing ED on a single cluster with open boundary conditions (OBC) embedded in a bath of self-consistently defined mean-fields \cite{isaev2009}.
In tensor-network language, the CGA Eq. \eqref{eqn:cga} is equivalent to a tree-tensor network with a single multi-qubit isometry with constraint $\sum_{\lbrace \sigma\rbrace} w^\ast_{\lbrace \sigma\rbrace} w_{\lbrace \sigma\rbrace}$=1  \cite{evenbly2009, tagliacozzo2009}.

The CGA energy \eqref{eqn:energy} on finite clusters provides an upper bound to the ground state energy of the model in its thermodynamic limit. Inspection of derivatives of the CGA energy unveils the phase diagram.
In addition, a finite-size scaling analysis allows the assessment of the stability of phases upon increasing the cluster size $N_c$ and allows extrapolation of the location of phase boundaries.
In this manner, the CGA provides a computationally inexpensive ansatz to approach models of frustrated quantum magnetism \cite{lacroix2011} that pose problems to state-of-the-art numerical approaches \cite{henelius2000,marvian2019}.

This simple yet expressive approximation has been applied to a variety of models where frustrated spin and bosonic interactions lead to the co-existence and competition of LRO and quantum paramagnetic phases, including valence-bond solids and chiral states \cite{isaev2009, isaev2009a, huerga2014, greschner2015, huerga2016}.
The algebraic framework of HMFT allows for other self-consistent mean-field approximations, including a Bogoliubov approximation that enables the study of low-lying excitations \cite{isaev2009} such as Goldstone and Higgs modes in superfluids \cite{huerga2013}. Moreover, HMFT can be extended to investigate finite-temperature  phase transitions \cite{isaev2012} and to construct parent Hamiltonians of valence-bond solids \cite{huerga2017}.

Here, we utilize clusters of size $N_c$=6 and 24 (see Fig. \ref{fig:phase_diagrams}), representing the two minimal instances preserving the C$_6$ rotational symmetry of the honeycomb lattice,
and systematically inspect the CGA energy and its derivatives to unveil the phase diagram of the KHM. 
We use the resulting CGA wavefunctions \eqref{eqn:cga} to compute observables characterizing the emergence of LRO, or lack thereof, within the phases thus obtained.
In addition we compute topological observables, such as the plaquette flux \eqref{eqn:wp} and the topological entanglement entropy \cite{kitaev2006a}.

\subsection{Exact diagonalization}
\begin{figure}[t]
\includegraphics[width=0.8\linewidth]{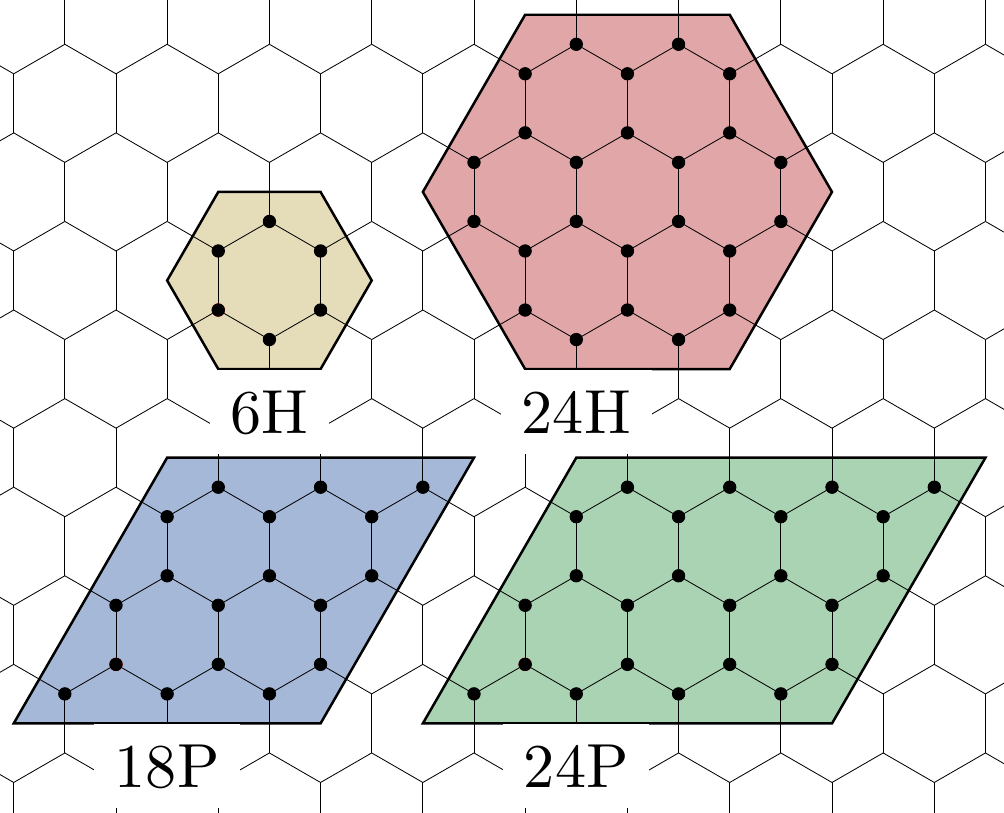}
\caption{Clusters used in exact-diagonalization with periodic boundary conditions. 6H and 24H clusters are identical to those used in HMFT calculations. 18P and 24P clusters are used for finite-size scaling analysis. }
\label{fig:ed_clusters}
\end{figure}
We make extensive use of exact-diagonalization (ED) in order to support at predictions arrived at through our HMFT simulations.
Specifically, we utilize the Lanczos method as implemented in the QuSpin package \cite{weinberg2017} to find energy and wavefunctions for the ground state and low-lying excitations using clusters of size $N_c$=18 and 24 with periodic boundary conditions (PBC) (see Fig. \ref{fig:ed_clusters}).
We use these results to obtain quantities that can indicate the emergence of QSLs or other topologically-ordered states, including the many-body Chern number \cite{ortiz1994,haldane1995,ortiz1996,fukui2005,varney2011} and the topological $S$-matrix \cite{zhang2012}. 

\section{Phase diagram\label{sec:results}}

\subsection{Benchmarking HMFT at  \texorpdfstring{$\boldsymbol{h=0}$}{h=0}}

To assess the validity of an HMFT description of QSL physics, we begin by studying the antiferromagnetic KHM given by Eq. \eqref{eqn:KHM} at $h$=0. This allows for direct comparison between HMFT results and those found via the model's exact solution \cite{kitaev2006}.

Using 6- and 24-site CGA, HMFT provides a quantitatively accurate approximation of the exact phase diagram.
In spite of the lack of a bulk in the 6-site cluster, HMFT on this cluster provides a qualitative picture of the boundaries between the A$_\gamma$ and KSL phases that becomes more accurate upon increasing the cluster size to $N_c$=24.
The quality of even the 6-site results are unsurprising given the very short correlation length characteristic of the pure KHM \cite{hermanns2018}.

In Fig. \ref{fig:phase_diagrams}(a), we show 6- and 24-site HMFT results on the quantum phase diagram of the KHM at $h$=0 along various cuts at fixed ratios of $K_y/K_x$.
By inspecting discontinuities in the derivatives of the energy \eqref{eqn:energy}, we identify a weakly first-order transition from A$_\gamma$ to KSL with 6-site HMFT that smoothes to second-order on the 24-site cluster, consistent with exact results (details are presented in Appendix \ref{app:h0}).
Plaquette flux \eqref{eqn:wp} computed with 6-site HMFT shows $W_p$=$-1$ for the intra-cluster plaquette, while the correct result $W_p$=$1$ is recovered in 24-site HMFT for \textit{all} intra-cluster plaquettes.\footnote{As the CGA wavefunction substitutes mean-fields for inter-cluster quantum correlations, $W_p$=0 for inter-cluster plaquettes.}
Additionally, topological entanglement entropy computed in the 24-site HMFT matches the exact solution  \cite{kitaev2006}, with $S_\text{topo}$$=$$-\log 2$ to within $\approx$$10^{-5}$ throughout the entire $h$=0 phase diagram. 

At $K_x$$=$$K_y$$=$$K_z$, we find a ground state degeneracy corresponding to different embedding mean-field configurations reflecting magnetic orders not found in the exact solution. 
Specifically, while the unique 6-site mean-field solution has all nearest neighbor spins aligned in opposite directions (N\'eel order), the 24-site cluster allows for four categories of mean-field configuration characterized by either N\'eel or stripe magnetic order and varying rotational symmetry. We refer to these as the C$_6$-stripy, C$_2$-stripy, C$_3$-N\'eel, and C$_2$-N\'eel configurations. Taking into account global rotations and sign flips, this leads to a total of 16 distinct configurations with identical energy.
\begin{figure*}[t!]    
\begin{tabular}{@{}c@{}}
    \includegraphics[width=0.2\textwidth]{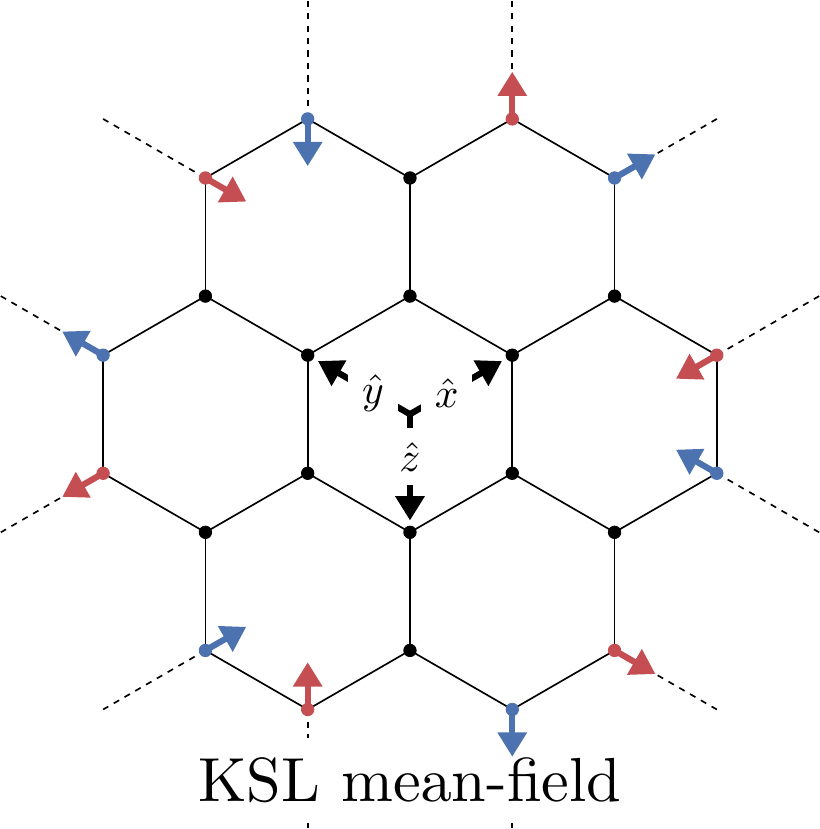}
    \end{tabular}
    $=$
    \begin{tabular}{@{}c@{}}
    \includegraphics[width=0.2\textwidth]{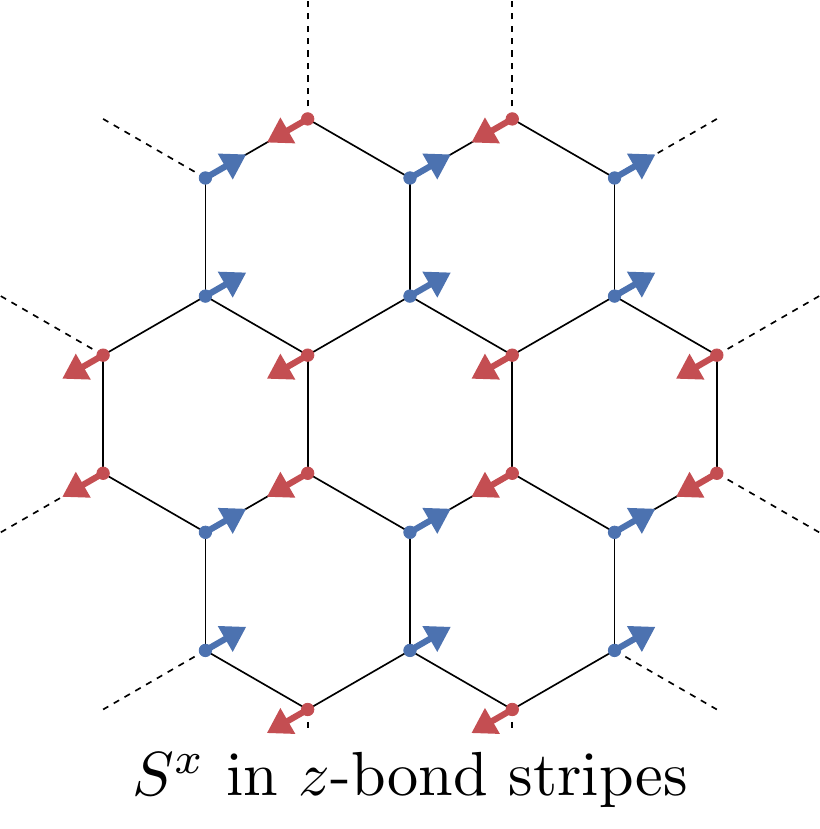}
    \end{tabular}
    $+$
    \begin{tabular}{@{}c@{}}
    \includegraphics[width=0.2\textwidth]{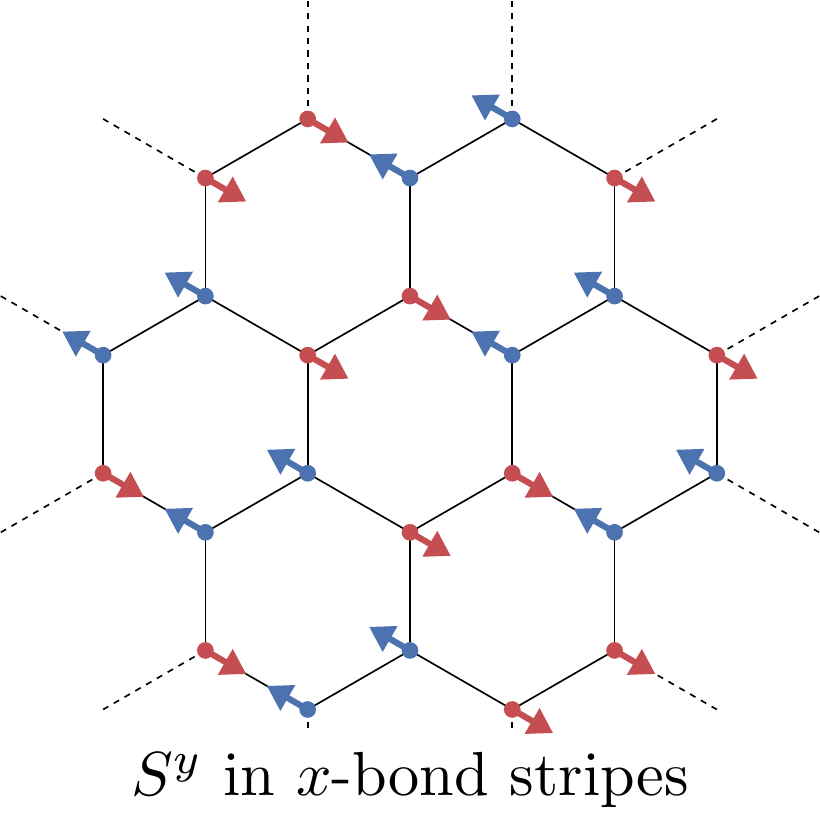}
    \end{tabular}
    $+$
    \begin{tabular}{@{}c@{}}
    \includegraphics[width=0.2\textwidth]{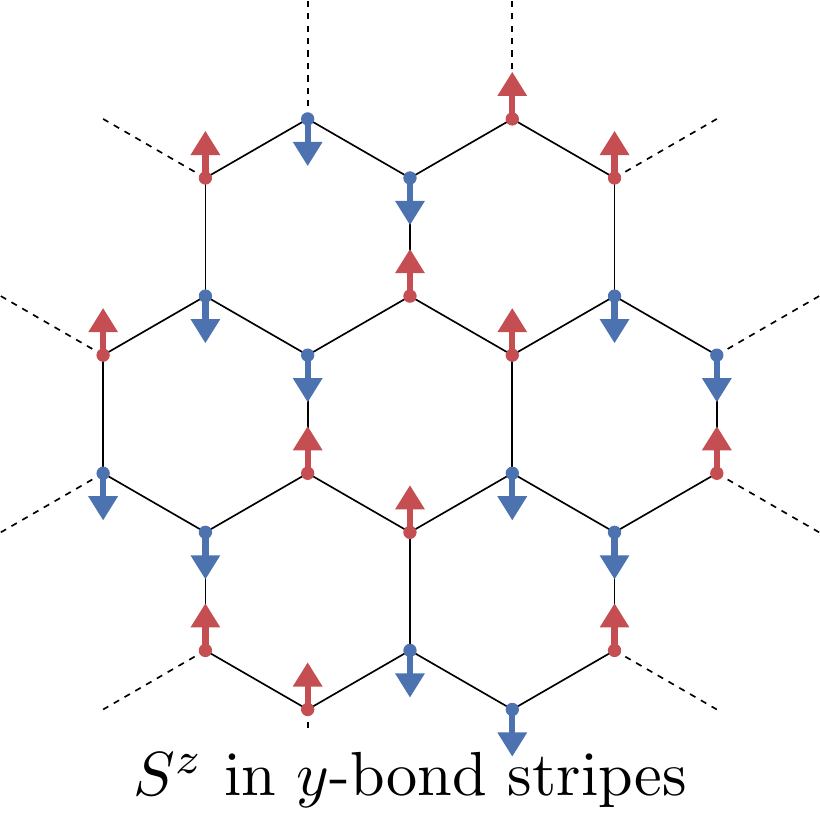}
    \end{tabular}
    \caption{Schematics of the C$_6$-stripy mean-field configuration describing the KSL phase as computed with 24-site HMFT. Blue (red) arrows represent positive (negative) spin along the [111] direction.}
    \label{fig:stripy_mf}
\end{figure*}
In Fig. \ref{fig:stripy_mf} we illustrate the C$_6$-stripy configuration, which is the only one preserving the C$_6$$\times$C$_3^S$ symmetry of the KHM at its maximally symmetric point.

Note that although LRO is generically concomitant to a non-zero mean-field embedding in CGA (Eq. \eqref{eqn:cga}) \cite{isaev2009,huerga2013,huerga2014}, in this case, spins located within the bulk of the 24-site cluster are completely paramagnetic, $\langle \mathbf{S}_i\rangle$=0. 
This causes the overall LRO signal to fall off as the ratio of cluster boundary to area with increasing cluster size, i.e. $\mathcal O(1/N_c)$. 
Therefore, the ``mean-field magnetic order'' appearing in the KSL is distinct from the LRO that we detect at $h$$\neq$0, to be discussed in the remainder of this work. 
As we will see, true LRO in HMFT is a property permeating the bulk of the cluster, and thereby persists in the $N_c\rightarrow\infty$ limit.
Further discussion of the different mean-field configurations is presented in Appendix \ref{app:order}.

\subsection{Kitaev honeycomb model in a [111] field}
\begin{figure}[t]
    \includegraphics[width=\linewidth]{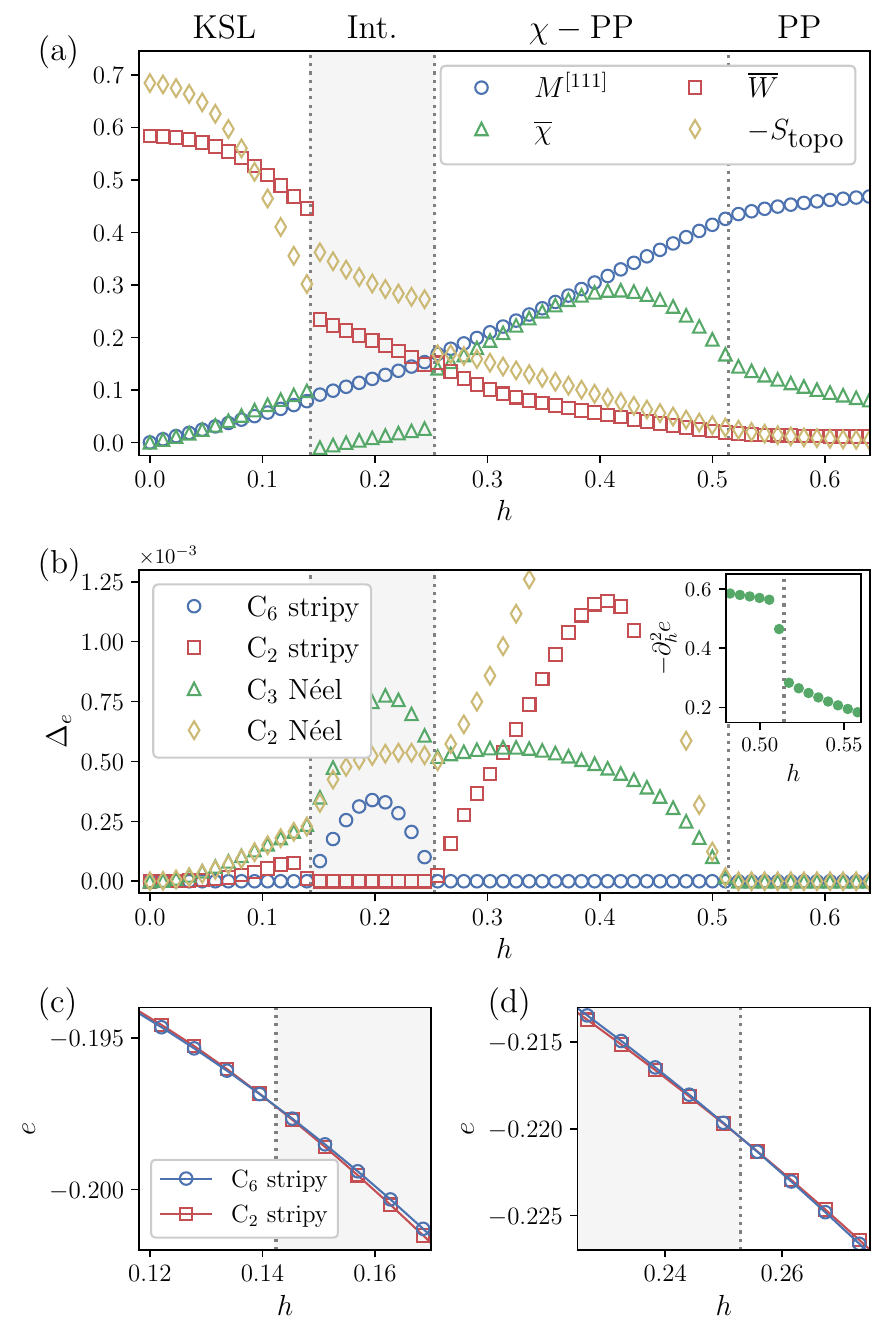}
    \caption{
    (a) Magnetization, chiral order parameter \eqref{eqn:chiral_op} and  average plaquette flux \eqref{eqn:wp-avg}, and topological entanglement entropy as computed with 24-site HMFT. 
    (b) Gaps of the lowest-lying competing HMFT solutions computed with respect to the optimal HMFT ground state energy. 
    Inset: second derivative of the energy showing a discontinuity at $h_3$.
    (c, d) Level crossings between the C$_6$- and C$_2$-stripy HMFT solutions at $h_1$ and $h_2$, respectively.
    }
    \label{fig:summary}
\end{figure}
We now focus on the finite field region $h$$>$0, for which no exact solution exists, and fix $K_x$=$K_y$=$K_z$=1 and $h$$\parallel$$[111]$, thus preserving the maximal symmetry of the KHM \eqref{eqn:KHM} and placing the ground state at $h$$\ll$1 deep within the the KSL phase.
To characterize magnetic order, we compute the magnetization along the $[111]$ direction, $M^{[111]}$=$1/(N_c\sqrt 3)\sum_{i,\gamma} \langle S^\gamma_i\rangle$
and the sublattice scalar chirality,
\beq
\chi_{ijk} = 2^3 \langle \mathbf{S}_i \cdot \left(\mathbf{S}_j \times \mathbf{S}_k \right)\rangle,
\label{eqn:scalar_chi}
\eeq
where $i,$ $j$ and $k$ are next-nearest neighbors on the honeycomb lattice.
Here and in all following equations, sums over the index $i$ are confined to the $N_c$ spins within a single cluster and all expectation values 
are taken with respect to the CGA wavefunction $\ket{\Psi}$ \eqref{eqn:cga} for HMFT calculations and the PBC ground state for ED calculations.
We define a global chiral observable as the average chirality over one sublattice, 
\beq
\overline{\chi}=\frac{1}{N_c}\sum_{\langle\langle ijk\rangle\rangle} w_{ijk}\chi_{ijk},
\label{eqn:chiral_op}
\eeq
where $\langle\langle ijk\rangle\rangle$ refers to next-to-nearest neighbor sites of the honeycomb lattice and weights take into account the cluster tiling of the lattice, i.e. $w_{ijk}$=1, 1/2, and 1/3, for $i,j,k$ belonging to one, two, or three clusters, respectively.
Due to the structure of the CGA wavefunction \eqref{eqn:cga}, the expectation value of an operator (i.e. $\chi_{ijk}$) acting on multiple clusters is equal to the product of expectation values taken within each cluster, meaning Eq. \eqref{eqn:chiral_op} can be evaluated using the wavefunction of a single cluster (see Appendix \ref{app:observables}).

We also compute the expectation value of the plaquette flux operator \eqref{eqn:wp} at every plaquette and define its average over the whole lattice,
\beq
\overline{W}=\frac{1}{N_c}\sum_p w_p \langle W_p\rangle,
\label{eqn:wp-avg}
\eeq
where, similarly to the chiral order parameter, the weight factors $w_p$ take into account whether the operator acts on one ($w_p$=1), two ($w_p$=1/2), or three ($w_p$=$1/3$) clusters.

In order to describe the topological character of QSLs, we compute the topological entanglement entropy $S_\textrm{topo}$ via the Kitaev-Preskill construction \cite{kitaev2006a} on the 24-site HMFT. 
Lack of a ``bulk'' (spins isolated from the cluster boundaries) in the 6-site cluster prevents computation of $S_\textrm{topo}$ with this cluster in HMFT. Details of this calculation are covered in Appendix \ref{app:topo}.

Figure \ref{fig:summary} illustrates our main results.
First, we find a low-field KSL phase adiabatically connected to the exact $h$=0 point, characterized by a positive average plaquette flux and topological entanglement entropy that decrease upon increasing $h$.
This KSL ends at a first order transition, leading to an intermediate phase exhibiting enhanced stripe magnetization along a preferred axis.
Before reaching the trivial partially-polarized (PP) phase with nearly-saturated [111] magnetization, we find a novel second intermediate phase characterized by the co-existence of finite scalar chirality \eqref{eqn:chiral_op} and partial polarization (thus $\chi$-PP).

Interestingly, the phase diagram can be broadly understood as two consecutive level crossings occurring between the C$_6$- and C$_2$-stripy solutions.
Specifically, the nonzero magnetic field breaks
the aforementioned degeneracy at $h$=$0$ in favor of the C$_6$-stripy solution within the KSL at finite $h$. 
At $h_1$$\approx$0.14, the C$_2$-stripy energy crosses below the C$_6$-stripy solution, becoming the new ground state of this intermediate phase and causing a first-order transition, albeit a subtle one due to the small difference in energy, as can be seen from the gap of order $10^{-3}$ in Fig. \ref{fig:summary}(b) and level crossings (Fig. \ref{fig:summary}(c) and (d)). 
At $h_2$$\approx$0.25, the situation reverses itself and the C$_6$-stripy solution crosses again, stabilizing the $\chi$-PP phase.
At $h_3$$\approx$0.51, we observe a continuous (second-order) phase transition towards the trivial PP phase signalled by a
large discontinuity in $\partial_h^2 e$.
At precisely this point, the C$_2$ and C$_6$-stripy solutions (along with the N\'eel-ordered solutions also degenerate at $h$=0) lose their distinction when their mean-field parameters converge to identical values.

\begin{figure}[t]
    \includegraphics[width=\linewidth]{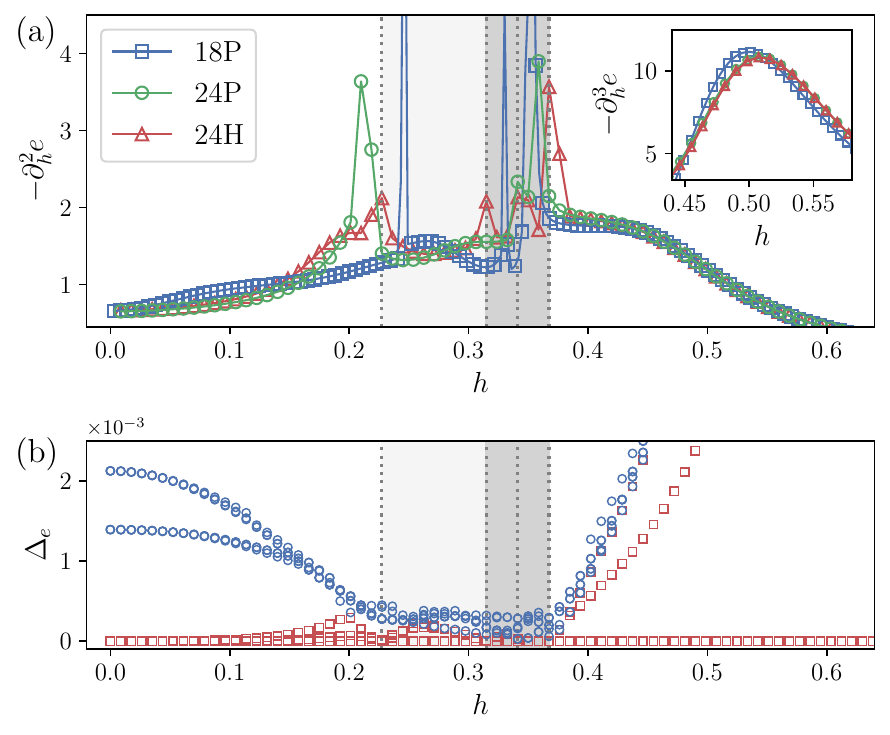}
    \caption{(a) Energy derivatives as computed with ED on hexagonal (H) and parallelogram (P) clusters with 18 and 24 sites using PBC. 
    Dashed lines indicate peaks in the second derivative from the 24H cluster. 
    The intermediate phase (shaded gray) comprises regions where the ground state is singly (light gray) and multiply degenerate (dark gray). 
    Inset: third derivative of the energy showing a smooth bump at $h$$\approx$0.5.
    (b) Gaps to low energy states. A 3-fold quasi-degeneracy can be distinguished within the KSL phase, with the the three lowest-energy states indicated by red squares.
    }
    \label{fig:ed_energy}
\end{figure}
Comparing these results with those obtained in previous ED \cite{zhu2018, hickey2019, ronquillo2019, pradhan2020} and DMRG \cite{zhu2018, gohlke2018, patel2019, pradhan2020} computations, 24-site HMFT predicts lower values of $h_1$ and $h_2$ and a transition at $h_3$ that has escaped previous numerical analysis.
In Fig. \ref{fig:ed_energy}, we show ED results from calculations performed on 18- and 24-site clusters (illustrated in Fig. \ref{fig:ed_clusters}) for comparison.
It can be seen that the value of $h_1$ obtained from ED decreases as the cluster size increases, moving towards the 24-site HMFT result.
Moving to $h_2$, ED shows a series of closely-spaced singularities in $\partial_h^2 e$ corresponding to a increase in ground-state degeneracy from the previously unique state, first to two-fold and then to a three-fold degeneracy.
The extent of this degenerate region decreases with increasing cluster size, suggesting that it corresponds to the single transition seen with HMFT and DMRG \cite{zhu2018, gohlke2018}.
Lastly, ED results on 18- and 24-site clusters show a peak in $\partial_h^3 e$ very close to the value of $h_3$ obtained from HMFT, but this peak is not accompanied by any other signature of a phase transition, including those typically appearing in the computation of fidelity susceptibility \cite{albuquerque2010}.

\subsubsection{Intermediate phase}

As mentioned previously, we find that the intermediate phase originates from a mean-field orientation with self-consistent fields that spontaneously break the C$_6$$\times$C$_3^S$ symmetry of the Hamiltonian in favor of a reduced C$_2$$\times$C$_1^S$ symmetry.
In order to characterize the spontaneous symmetry breaking (SSB) found in the intermediate phase, we inspect 
stripe-order staggered magnetization,
\beq
M^\gamma_{\alpha-\text{stripe}}=\frac{1}{N_c}\sum_{i}s_i \braket{S_i^\gamma},
\label{eqn:stripe}
\eeq 
where $\alpha$=$x,y,z$ refers to the bond direction along which nearest neighbors are aligned
and $s_i$$=$$\pm1$ depending on which of the to two sets of stripes (with opposed spins) perpendicular to the $\alpha$ bonds site $i$ belongs to (Fig. \ref{fig:stripy_mf}).
In addition, to directly indicate SSB, we define an onsite observable
\begin{equation}
\mathcal O_i = 
\left|\braket{\mathbf{S}_i - \hat{\mathbf{U}}^{-1} \mathbf{S}_i \hat{\mathbf{U}}}\right|,
\label{eqn:ssb_op}
\end{equation}
where $\hat{\mathbf{U}}$ is a unitary operator implementing a C$_6$$\times$C$_3^S$ rotation.
If $\mathcal O_i$$=$$0$, the system is symmetric. Otherwise, $\mathcal O_i$$>$$0$ signals broken C$_6$$\times$C$_3^S$ symmetry. 

\begin{figure}
    \includegraphics[width=\linewidth]{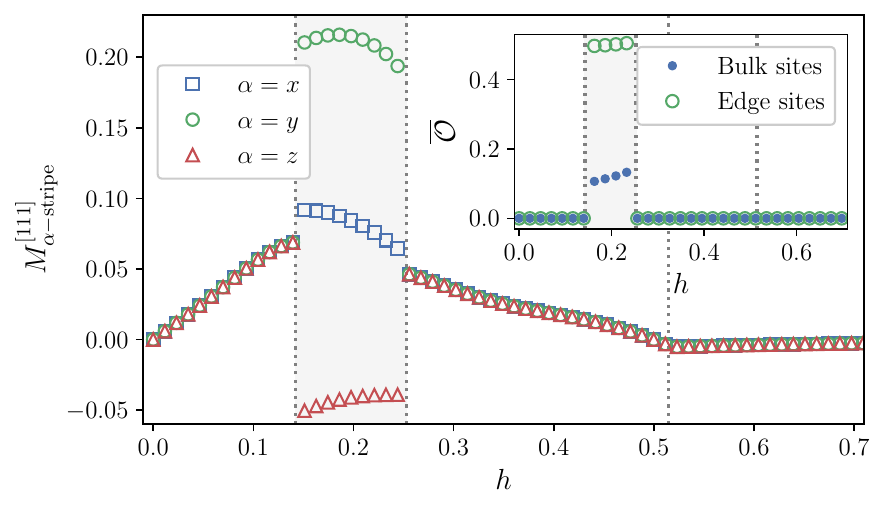}
    \caption{Observables capturing the spontaneous symmetry breaking (SSB) exhibited by the intermediate phase obtained from 24-site HMFT. Main: stripe magnetizations measured along the $[111]$ spin direction for stripes perpendicular to $x$, $y$, and $z$ bonds. Inset: symmetry-breaking parameter \eqref{eqn:ssb_op} averaged over sites in the cluster's bulk and boundary.}
	\label{fig:stripe}
\end{figure}
In Fig. \ref{fig:stripe} we show stripe magnetization along the [111] direction, $M_{\alpha-\text{stripe}}^{[111]}$=$\frac{1}{\sqrt 3}\sum_\gamma M_{\alpha-\text{stripe}}^\gamma$,
together with the observable \eqref{eqn:ssb_op} averaged over the cluster's bulk and boundary, as computed with 24-site HMFT.
We find the magnetization depends on stripe direction $\gamma$ only in the intermediate phase, with $\gamma$-independent values in the KSL, $\chi$-PP, and PP phases.
In particular, $M_{y-\textrm{stripe}}^{[111]}$ increases within the intermediate phase,\footnote{The preference towards $M_{y-\textrm{stripe}}^{[111]}$ in particular is arbitrary, and only reflects the orientation our mean-fields selected. Rotations of these fields result in an equivalent HMFT state with the same energy and a different preferred stripe orientation.} exceeding $20\%$ of saturation, while the others stripe magnetizations also increase in magnitude from their values in other phases.
The dependence of stripe magnetization on $\gamma$ already establishes SSB in the intermediate phase, but its presence is further supported by the nonzero values of $\mathcal O_i$ found in the intermediate phase (inset of Fig. \ref{fig:stripe}). The larger value of $\mathcal O_i$ on cluster boundaries is due to the HMFT fields, which drive the SSB. Unlike the strictly mean-field order found at $h$=0, bulk spins also acquire magnetic order (hence nonzero $\mathcal O_i$) in the intermediate phase.

As a consequence of this lack of SSB in finite systems, the stripe-order magnetizations \eqref{eqn:stripe} measured on the symmetric 24H cluster with PBC (Fig. \ref{fig:ed_clusters}) have zero expectation value. 
To look for signatures of symmetry-breaking LRO in ED, we compute the staggered-field susceptibility,
\beq
\chi_{\alpha\textrm{-stripe}}^{\beta\gamma}  =
\partial_\varepsilon \left. \braket{\Psi_{\alpha}^\beta(\varepsilon)|\hat{M}_{\alpha\textrm{-stripe}}^\gamma|\Psi_{\alpha}^\beta(\varepsilon)}\right|_{\varepsilon= 0},
\label{eqn:susc}
\eeq
where $\ket{\Psi_{\alpha}^\beta(\varepsilon)}$ is the ground state of the perturbed Hamiltonian, $H(\varepsilon)$=$H$$+$$\varepsilon \hat{M}_{\alpha\textrm{-stripe}}^\beta$,
and examine the static spin structure factor
\beq
S(\mathbf k) = \frac{1}{N_c^2}\sum_{i,j}e^{i\mathbf r_{ij}\mathbf k} \braket{\mathbf S_i \mathbf S_j} - 
\frac{1}{N_c}\left|\sum_i e^{i\mathbf r_i \cdot \mathbf k}\braket{\mathbf S_i}\right|^2,
\label{eqn:struc_fac}
\eeq
where $\mathbf r_{ij}$=$\mathbf r_i - \mathbf r_j$, 
at various $\mathbf k$-points commensurate with the 24-site honeycomb cluster (24H) and representative of various types of magnetic LRO. 
In particular, the points $M$, $M_e$, and $\Gamma_e$ correspond to 
zig-zag, stripe, and N\'eel staggered magnetizations, respectively (see Appendix \ref{app:kspace}). 

\begin{figure}[t]
    \includegraphics[width=\linewidth]{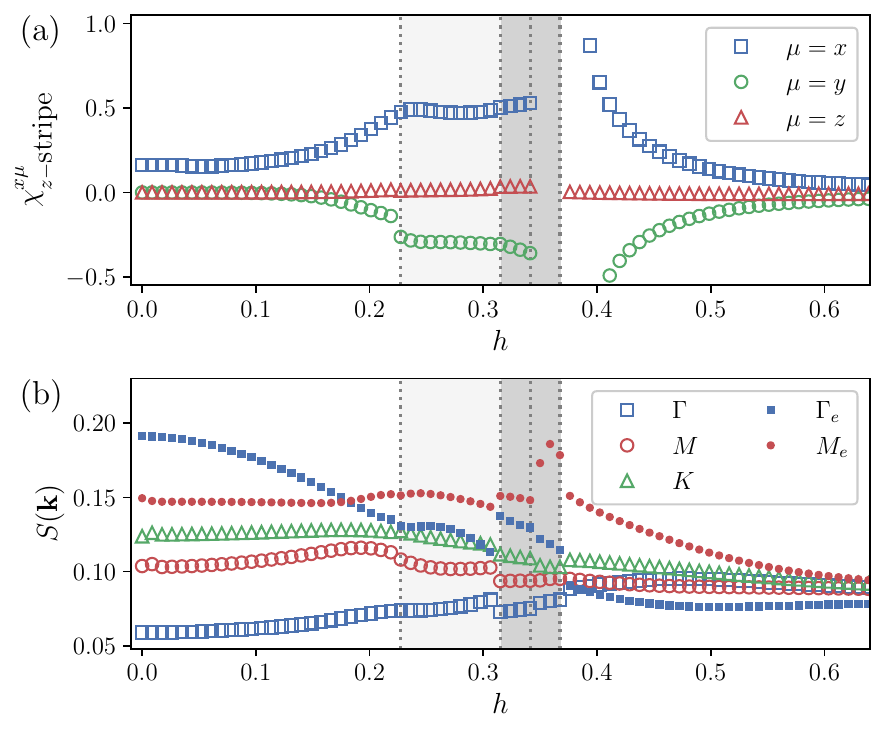}
    \caption{(a) Staggered magnetic susceptibility \eqref{eqn:susc}, and (b) static structure factor \eqref{eqn:struc_fac} for different high-symmetry points of the Brillouin zone, as computed with ED on 24H cluster with PBC.}
	\label{fig:ed_stripe}
\end{figure}
In Fig. \ref{fig:ed_stripe}(a) we observe an increase in the magnitude of susceptibilities $\chi_{z\textrm{-stripe}}^{xx}$ and $\chi_{z\textrm{-stripe}}^{xy}$ in the intermediate phase, while $\chi_{z\textrm{-stripe}}^{xz}$ is near zero for all $h$, consistent with the presence of antiferromagnetic $\sum_{\langle ij\rangle^z}S^z_i S^z_j$ interactions in the Hamiltonian on precisely those bonds that  $z$-direction stripes would align.
In addition to the susceptibility, the static structure factor at $M_e$, which corresponds to stripe order, is larger than at any other $\mathbf{k}$-point within the intermediate phase, as can be seen in Fig. \ref{fig:ed_stripe}(b), consistent with previous DMRG calculations that identified an enhanced signal in this same phase \cite{patel2019}.

In addition to ruling out a featureless QSL, the LRO we find in the intermediate phase also complicates predictions of a topological $C$=$4$ state arrived at from effective mean-field \cite{zhang2021} and variational Monte Carlo studies \cite{jiang2020}.
In the interest of further exploring this, we probe the topological nature of the intermediate phase by measuring the many-body Chern number \cite{ortiz1994,haldane1995,fukui2005,varney2011} on the 24H cluster in ED.
This is a highly involved computation that requires integrating over a discretized torus $L$$\times$$L$ of twisted boundary conditions (TBC) \cite{haldane1995} (see Appendix \ref{app:topo} for details). 
Our computations for several TBC torus grids ($L$=6, 8, 10, 12) indicate that the KSL and PP phases are characterized by $C$=1 and $C$=0, respectively, consistent with the exact limits (KSL at $h$=0, and PP at $K_\gamma$=0).
However, we find the Chern number to be ill-defined within the intermediate phase, where $C$ jumps between different integer values throughout the single ground state region, and cannot be defined for the regions where the ground state is multiply degenerate (shaded dark gray in Fig. \eqref{fig:ed_energy}(b)). 
In addition, the ground state degeneracies we find are inconsistent with a $C$=4 state per Kitaev's 16-fold classification, which requires four-fold degeneracy or quasi-degeneracy \cite{kitaev2006}.
In our judgement, the lack of a well-defined Chern number most likely indicates a gapless spectrum, in agreement with previous predictions \cite{gohlke2018, hickey2019, patel2019}.

\subsubsection{Chiral partially-polarized phase}
\begin{figure}
    \includegraphics[width=\linewidth]{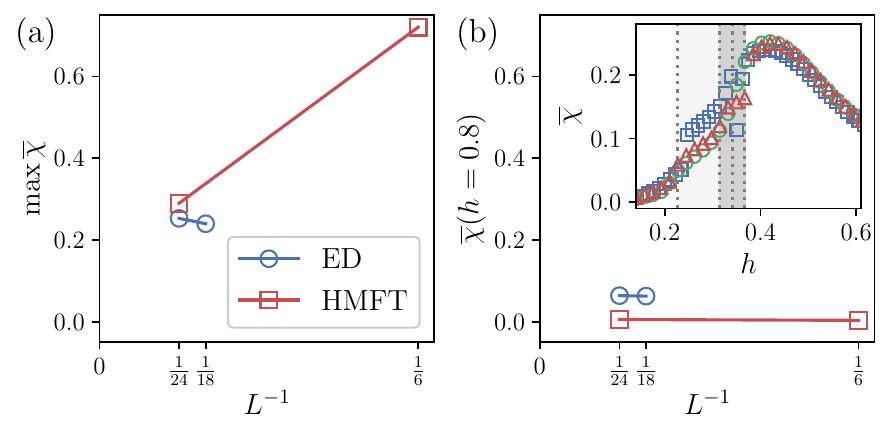}
    \caption{
    (a) Finite-size scaling of maximum chirality in the $\chi$-PP phase.  
    ED results shown here were obtained from 18P and 24P clusters (Fig. \ref{fig:ed_clusters}) using PBC.
    (b) Finite-size scaling of chirality in the PP phase at $h$=$0.8$.
    Inset: scalar chirality measured on 18- and 24-site clusters with PBC. Here, squares, triangles, and circles correspond to 18P, 24P, and 24H clusters.
    }
    \label{fig:ed_chi}
\end{figure}
The combined use of spatially symmetric clusters, together with the self-consistent mean-field embedding providing information from the thermodynamic limit and explicitly allowing for the breakdown of continuous symmetries, permits HMFT to discover phase transitions that may escape other methods.
That is the case of the second-order phase transition we observe at $h_3$$\approx$0.51, which separates a previously unnoticed chiral region from the trivial partially-polarized (PP) phase.
This $\chi$-PP phase is characterized by co-existence of partial polarization, $M^{[111]}$, and a large sublattice chirality \eqref{eqn:scalar_chi}, as illustrated in Fig. \ref{fig:summary}.

In Fig. \ref{fig:ed_chi} we perform a finite size scaling of the sublattice chirality,
as computed with HMFT and ED.
Within the $\chi$-PP phase, the maximum sublattice chirality obtained from both methods extrapolate to a similar finite value, $\overline{\chi}$$\approx$0.25, with HMFT approaching from above and ED from below. 
In the PP phase, both ED and HMFT show a strongly suppressed signal. 
Interestingly, the average chirality as computed with ED (plotted in the inset) exhibits a dependence on $h$ close that found in 24-site HMFT, with a maximum value at almost the same magnetic field in both 24-site ED and 18- and 24-site ED, $h$$\approx$0.4.

\begin{figure}
    \includegraphics[width=\linewidth]{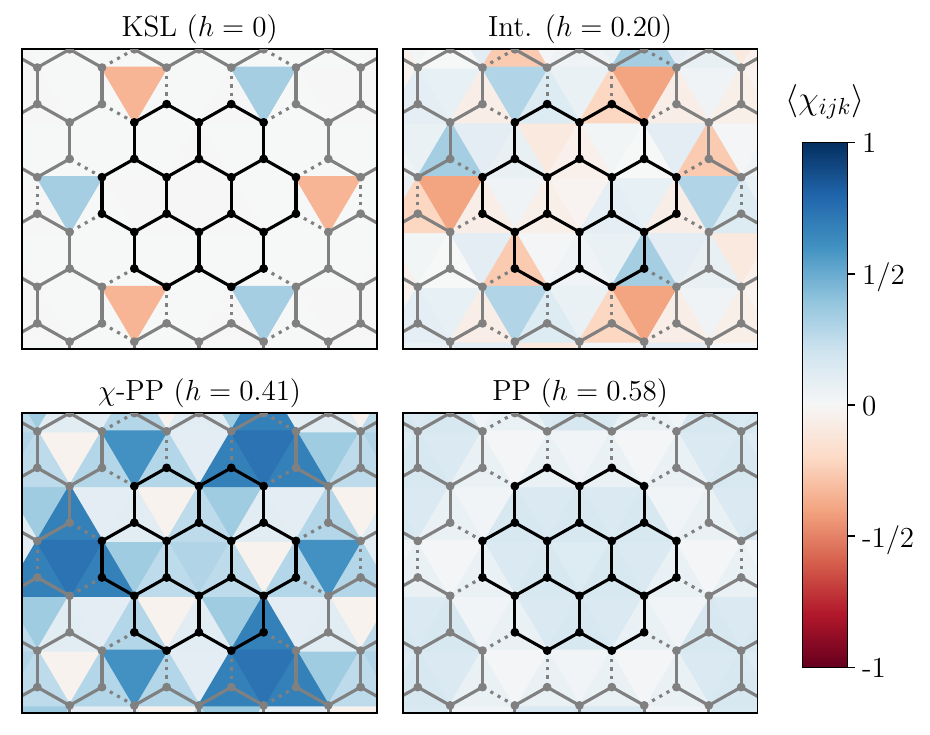}
    \caption{
    Scalar chirality distribution within the cluster at magnetic fields representative of the phases seen at $h$$>$0.}
    \label{fig:chi_heatmap}
\end{figure}

In Fig. \ref{fig:chi_heatmap}, we show that local sublattice scalar chirality
permeates the cluster within the $\chi$-PP phase, while in the other two non-trivial phases (KSL and intermediate) its effect is mostly present at the boundaries of the cluster.

\section{Discussion and conclusion}
\label{sec:conclusion}
Indications of field-revealed quantum spin liquid (QSL) behavior and topological order make frustrated magnets in the presence of external magnetic fields a subject of experimental and theoretical research attracting much attention.
The exactly-solvable Kitaev honeycomb model (KHM) is an important model of topological QSL physics  \cite{kitaev2006} and has motivated the search for its material realization since its proposal \cite{jackeli2009, chaloupka2010}, including extensive theoretical studies of the resulting extended Kitaev models \cite{janssen2016, gordon2019, yang2020, sorensons2021, yang2021, zhang2021a}.

In this work, we have approached the antiferromagnetic KHM by means of hierarchical mean-field theory (HMFT) supplemented with exact diagonalization (ED), finding that a magnetic field drives the exactly solvable Kitaev spin liquid (KSL) phase through two intermediate phases characterized by stripe and chiral magnetic orders, respectively, before transitioning into a trivial partially-polarized phase.
The first intermediate phase is characterized by spontaneous symmetry breaking and onset of stripe magnetic order, contrary to previous results arguing for a U(1) gapless \cite{zhu2018, gohlke2018, hickey2019, patel2019, ronquillo2019, pradhan2020} or gapped topological \cite{zhang2021, jiang2020} QSL.
This is supported by susceptibility, static spin structure factor, and many-body Chern number computed from ED that, taken together, provide strong evidence for the emergence of stripe order and, more specifically, rule out the possibility of a gapped topological ground state in this phase. 

The second intermediate phase, the chiral partially polarized ($\chi$-PP), is characterized by emergence of a clear signal of sublattice chirality co-existing with partial polarization.
This enhanced chirality is also observed in ED results across multiple clusters, with finite-size scaling of both HMFT and ED results indicating that the chiral order persists into the thermodynamic limit. 
Interestingly, this chiral phase is characterized by many-body Chern number $C$=0. Contrary to a commonly held belief, such a Chern number can be zero in a chiral phase, as it is a unique measure of the topology of the many-body wavefunction \cite{waldtmann1998}.
It is also worth noting that previous studies on quasi-one-dimensional versions of the KHM have also argued for the presence of chirality in the surrounding phase space \cite{sorensons2021}.

Despite showing indications of chiral order with remarkable similarity to those seen in HMFT, ED results do not exhibit any signatures of a phase transition separating it from the trivial partially polarized (PP) phase. 
Although a peak is present in the third derivative of the energy at similar magnetic fields where the transition is found in HMFT, this peak is not accompanied by a gap closing.
It is therefore possible that either the observed phase transition between the $\chi$-PP and PP phases is an artifact of HMFT and the two phases are in fact adiabatically connected or that its absence in ED is simply due to small system sizes.
Nevertheless, the robust scalar sublattice chirality observed in both ED and HMFT, together with the fact that the chirality permeates the bulk of the HMFT cluster only within the $\chi$-PP phase, indicates scalar chirality plays an important role in the physics of the Kitaev model in a range of magnetic fields above the first intermediate phase.

HMFT provides us with a broad picture of the phase diagram that can be understood as two consecutive crossing between two HMFT solutions, opening the intermediate phase, and a second-order phase transition separating the chiral and partially polarized phases, at which point these solutions become equivalent.

It is instructive to consider why our results, especially regarding SSB in the intermediate phase, were not seen in ED and DMRG studies. 
Unlike these methods, HMFT simultaneously simulates the thermodynamic limit (as in infinite-DMRG) and preserves two-dimensional symmetries of the model (as is possible in ED). 
Without meeting both conditions, the SSB we predict cannot be directly observed.

To rigorously confirm the ultimate fate in the thermodynamic limit of the chiral and stripe orders predicted in this study, clusters of sizes greater than $N_c$=24 may be required. 
This forms a bottleneck for classical computational methods. Instead, novel approaches may be required to approach larger clusters, which may utilize entanglement renormalization ideas \cite{vidal2007}, Monte Carlo methods \cite{honecker2016,alet2016}, the use of quantum computational resources \cite{huerga2022}, or combinations thereof.

{\bf Acknowledgments.}
This research was undertaken thanks in part to funding from the Canada First Research Excellence Fund. 

\newpage

\bibliographystyle{ieeetr}

\clearpage

\appendix

\section{\texorpdfstring{$\boldsymbol{h=0}$}{h=0} HMFT results}
\label{app:h0}

To determine the HMFT phase boundaries of the Kitaev honeycomb model (KHM) at $h$=0, we performed HMFT calculations iteratively (reusing the previous iteration's mean-fields as starting parameters) moving along paths originating at the $K_x$$=$$K_z$ line and ending at the $K_z$=1, $K_x$$=$$K_y$$=$$0$ point, with the ratio $K_x/K_y$ fixed along the path. These paths are illustrated in Fig. \ref{fig:phase_diagrams}(a), along with the phase boundaries of the KHM from the exact solution.

\begin{figure}
    \includegraphics[width=\linewidth]{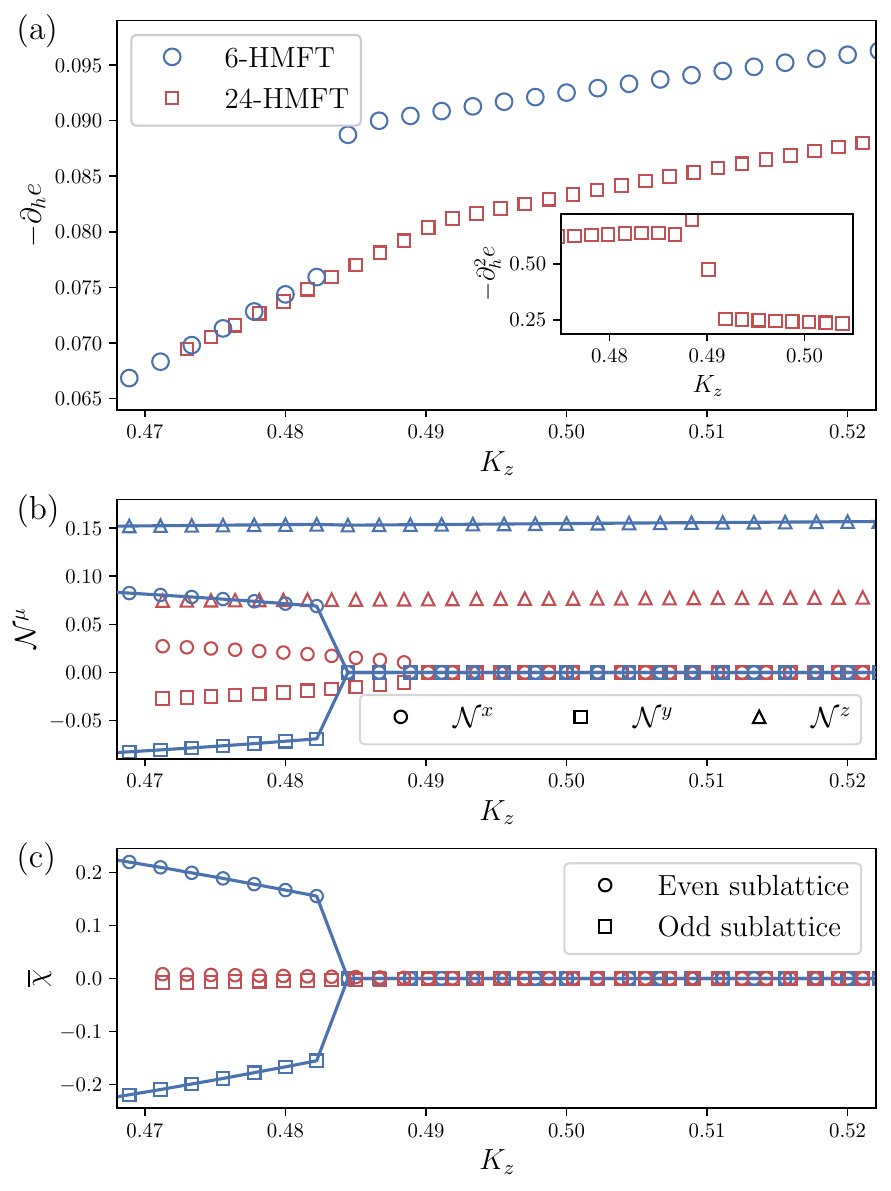}
    \caption{Results calculated at $h$=0 near the A$_z$-KSL phase transition calculated along the $K_x=K_y$ line.
    Blue and red symbols correspond to 6- and 24-site HMFT, respectively. 24-site results are obtained from a C$_3$-symmetric N\'eel-ordered mean-field.
    (a) First derivative of energy density showing a discontinuity in the 6-site results and a cusp in the 24-site results. Inset: second derivative of the 24-site energy density showing a discontinuity at the transition.
    (b) N\'eel order parameter measured along $x$, $y$, and $z$ directions.
    (c) Scalar chirality $\chi$ averaged over even and odd-sublattice triangles.}
    \label{fig:h0_results}
\end{figure}

Figure \ref{fig:h0_results} illustrates key results from the $h$=0 calculations calculated along the $K_x$$=$$K_y$ line (vertical in Fig. \ref{fig:phase_diagrams}(a)) indicating the transition between KSL and A$_z$ phases, occurring at $K_z$$=$$0.5$ in the exact solution. First derivatives of the energy show a first-order transition in the 6-site HMFT results, while 24-site HMFT correctly recovers a second-order transition, with only a cusp in the first derivative (Fig. \ref{fig:h0_results}(a)).

Due to the HMFT mean-fields, both N\'eel and chiral order are apparent in the HMFT solutions. The N\'eel order plotted in subfigure (b) of Fig. \ref{fig:h0_results} is illustrative of the mean-field structure: in the KSL phase, N\'eel order exists along all spin directions. At precisely the transition into the $K_z$-dominated A$_z$ phase, $\mathcal N^x$ and $\mathcal N^y$ go to zero, leaving only $z$-direction N\'eel order. This is much more visible in the 6-site results than the 24-site. Examination of the spatial dependence of spin expectation values shows that the N\'eel order in the 24-site cluster is only present in the boundaries (sites directly coupled to mean-fields). As such, it is likely that N\'eel order would disappear roughly as the ratio of boundary to area of the cluster (so $\mathcal O\l(1/N_c\r)$) for even larger cluster sizes ($N_c$$>$$24$). Note that this simple scaling does not apply to situations such as chirality in the $\chi$-PP phase, where significant magnetic order exists not only at the boundaries, but within the clusters as well.

Subfigure (c) of Fig. \ref{fig:h0_results} shows scalar chirality averaged over all triangles in the clusters. In the KSL phase, the even and odd sublattices acquire chirality in opposite directions, which goes to zero in the A$_z$ phase. Again, the magnitude of chirality is much smaller in the 24-site results. In fact, the only triangles with nonzero chirality at $h$=0 are those linking three clusters (see Fig. \ref{fig:chi_heatmap}). As with N\'eel order, we expect this indicates a strong decrease in chirality with further increases in cluster size.

Here, it should be noted that we computed observables in Fig. \ref{fig:h0_results} using the C$_3$-symmetric N\'eel-ordered 24-site HMFT configuration to simplify comparison to the single (C$_3$-symmetric N\'eel-ordered) 6-site HMFT solution. The stripe-ordered solutions relevant at $h$$>$0 (and degenerate at $h$=0) replace N\'eel order with stripe ordering (N\'eel-ordered solutions have precisely zero stripe magnetization and visa-versa).
Additionally, average chirality $\overline \chi$ is zero in the stripe-ordered solutions, with a pattern of positive and negative chiralities throughout the cluster (see Fig. \ref{fig:chi_heatmap}) resulting in an exact cancellation when summed for all couplings.

\section{Mean-field orientations}
\label{app:order}
\begin{figure}
    \begin{subfigure}[b]{0.325\linewidth}
    \includegraphics[width=\textwidth]{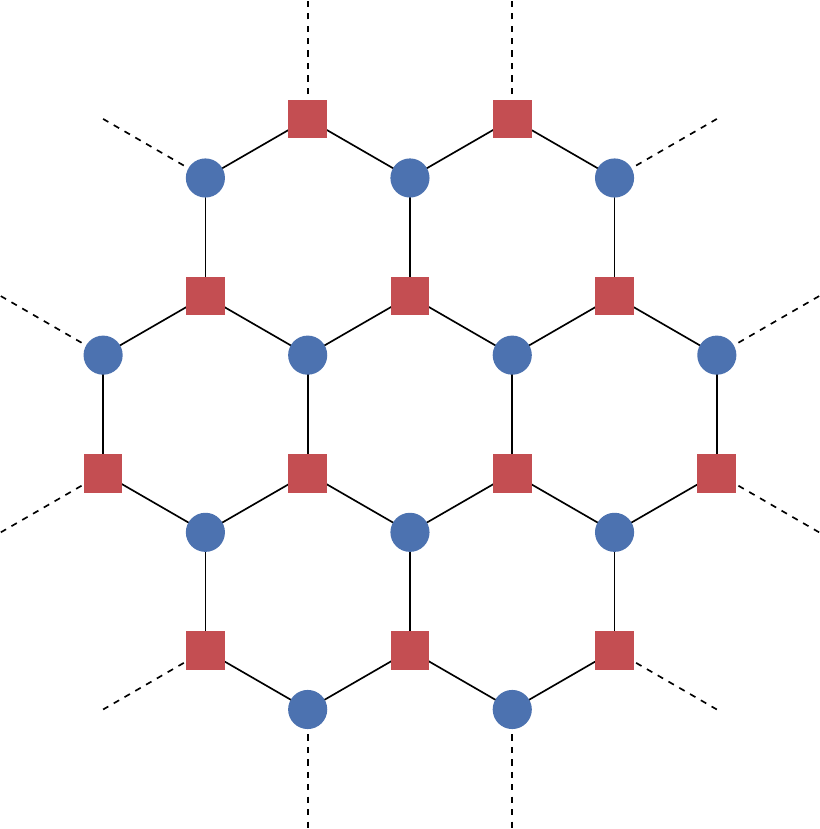}
    \caption{}
    \end{subfigure}
    \hfill
    \begin{subfigure}[b]{0.325\linewidth}
    \includegraphics[width=\textwidth]{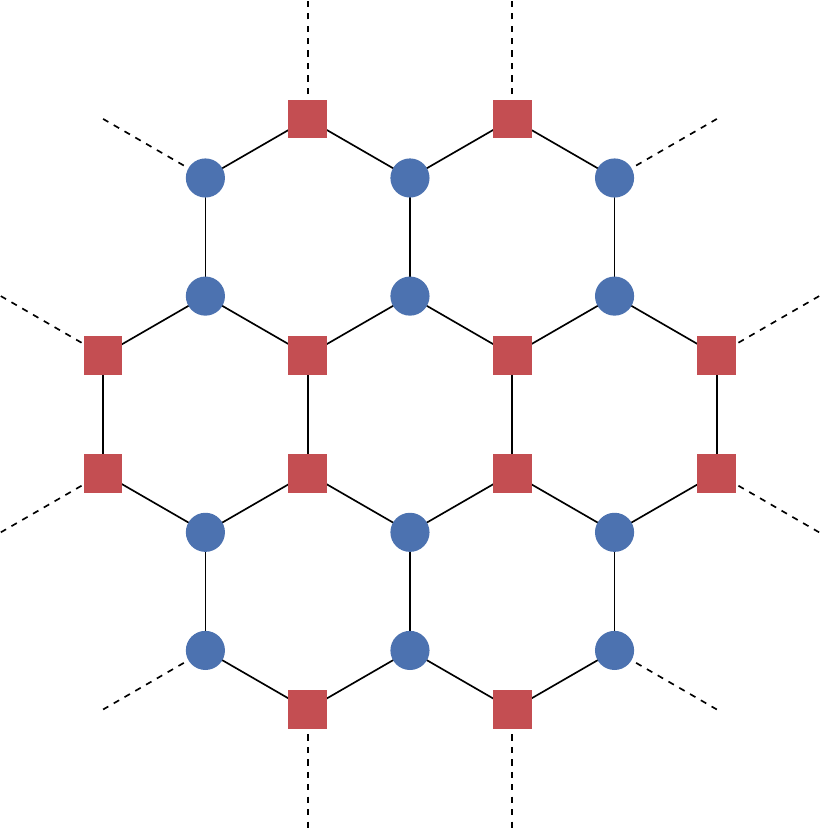}
    \caption{}
    \end{subfigure}
    \hfill
    \begin{subfigure}[b]{0.325\linewidth}
    \includegraphics[width=\textwidth]{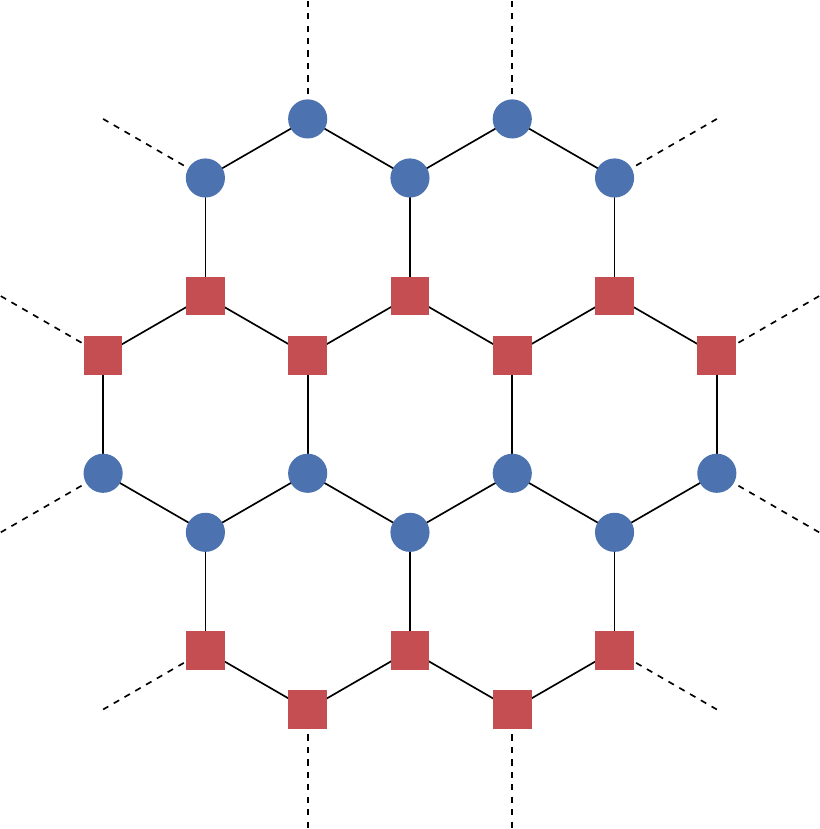}
    \caption{}
    \end{subfigure}
    \caption{N\'eel (a), stripe (b), and zig-zag (c) staggered magnetic orders. Rotations of the stripe and zig-zag orders give three orientations each.}
    \label{fig:af_order}
\end{figure}

Various staggered magnetizations occuring on the honeycomb lattice are shown in Fig. \ref{fig:af_order}. Note that the 24-site cluster is commensurate with all three orderings, while the 6-site cluster is only commensurate with N\'eel order.
We find self-consistent mean-fields with nonzero N\'eel and stripe order in the 24-site cluster, while zig-zag ordering is not seen for the antiferromagnetic interactions used in our simulations. Note that applying a spin flip to all even sublattice spins in the stripe-ordered arrangement transforms it to the zig-zag arrangement and visa versa. This indicates that the degeneracy between N\'eel and stripe order in the antiferromagnetic KHM HMFT solution corresponds to a degeneracy between zig-zag and uniformly magnetized mean-fields in the ferromagnetic KHM, as the models are identical up to the same sublattice spin flip.

\begin{figure}
    \begin{subfigure}[b]{0.48\linewidth}
    \includegraphics[width=\textwidth]{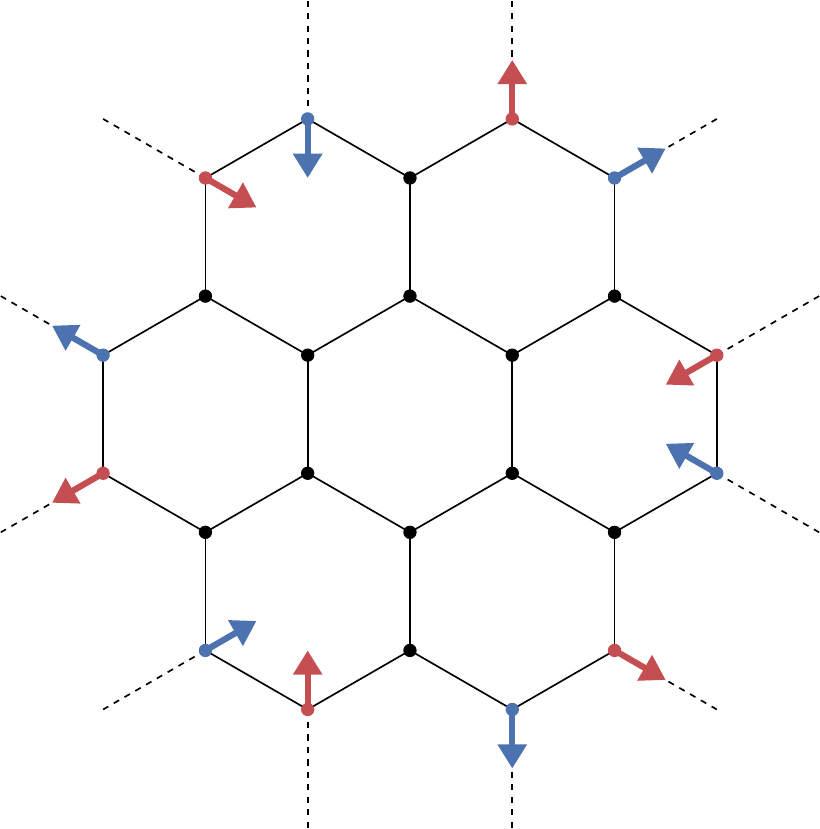}
    \caption{C$_6$-stripy}
    \end{subfigure}
    \hfill
    \begin{subfigure}[b]{0.48\linewidth}
    \includegraphics[width=\textwidth]{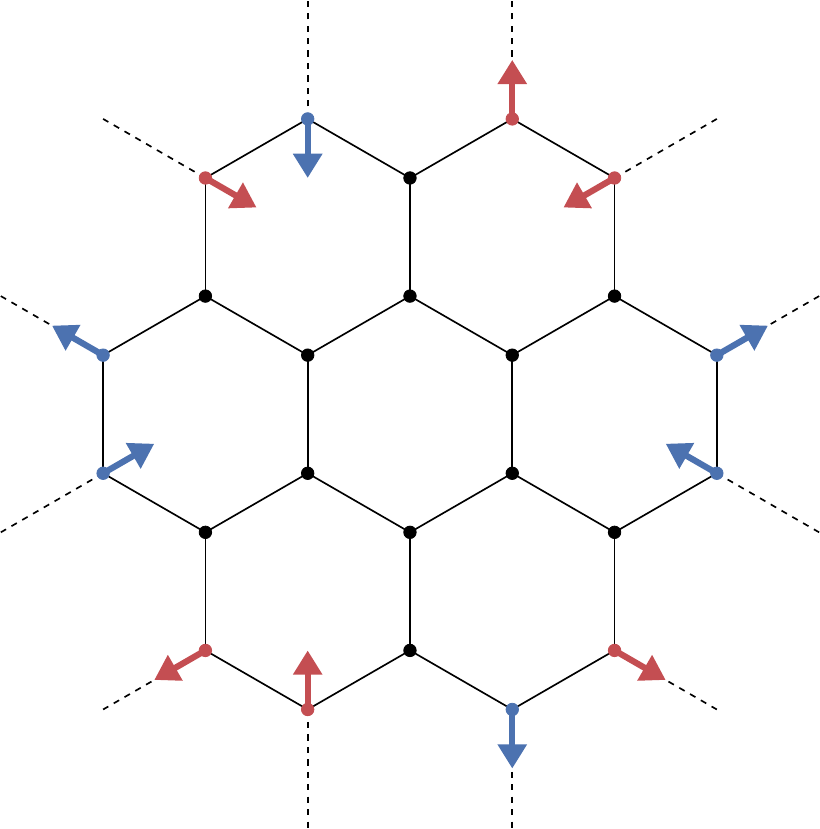}
    \caption{C$_2$-stripy}
    \end{subfigure}
    \hfill
	\begin{subfigure}[b]{0.48\linewidth}
    \includegraphics[width=\textwidth]{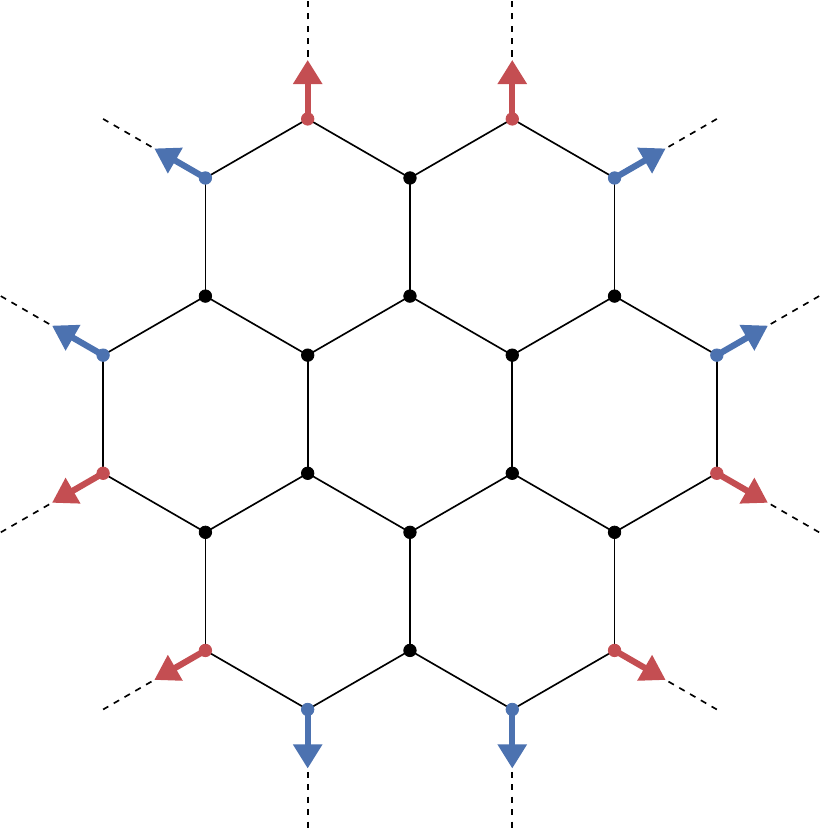}
    \caption{C$_3$-N\'eel}
    \end{subfigure}
    \hfill
    \begin{subfigure}[b]{0.48\linewidth}
    \includegraphics[width=\textwidth]{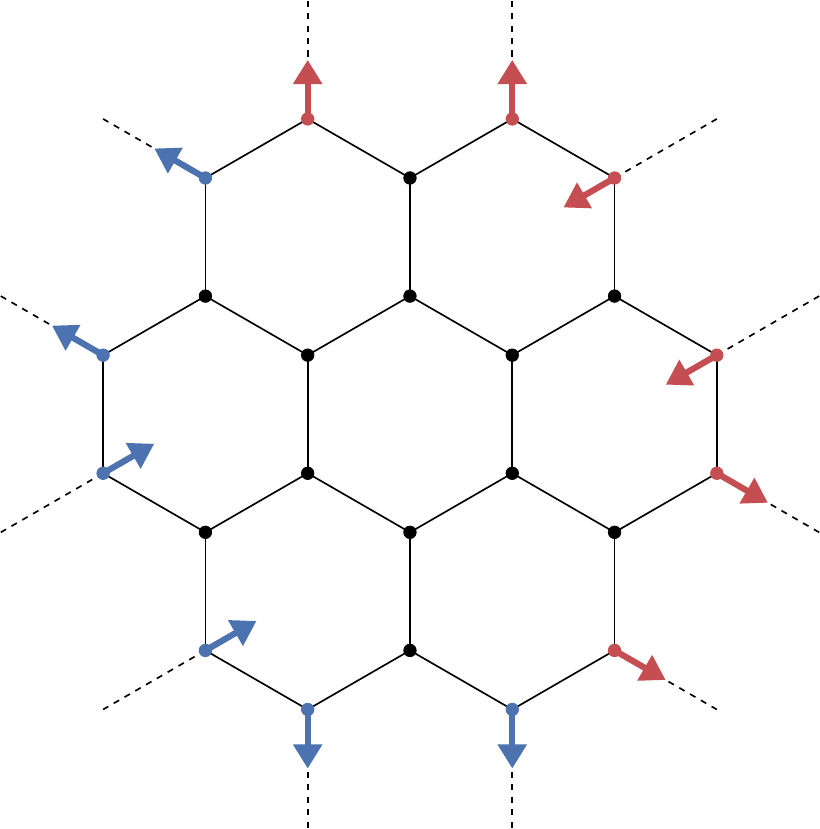}
    \caption{C$_2$-N\'eel}
    \end{subfigure}
    
    \caption{Mean-field orientations degenerate at $h$=0 and $K_x$$=$$K_y$$=$$K_z$. Here, spins are projected in the directions shown in (a), with red (blue) indicating positive (negative) sign.}
    \label{fig:mf_order}
\end{figure}

Figure \ref{fig:mf_order} shows the four categories of mean-field configurations with identical energy when $K_x$$=$$K_y$$=$$K_z$$=$$1$ and $h$=0. While the C$_3$-symmetric N\'eel order shown conforms to familiar N\'eel order where all sites on the even sublattice have $\braket{S_i^\mu}$ with an opposite sign to those on the odd sublattice, the other orderings are more correctly thought of as being commensurate with the labelled antiferromagnetic orders, rather than being a direct example for them. For instance, only the $x$-bond mean-fields in the C$_6$-symmetric stripe orientation shown in Fig. \ref{fig:mf_order} are consistent with the stripe orientation shown in Fig. \ref{fig:af_order}. The $y$ and $z$-bond mean-fields are consistent with other stripe orientations. As such, these mean-field orders represent overlayed staggered magnetizations with a different ordering for each component of spin.

Order parameters for the staggered magnetizations shown in Fig. \ref{fig:af_order} can be constructed as
\beq
M_{\textrm{staggered}}^\mu = \frac{1}{N_c}\left(\sum_{i\in\textrm{ red sites}}\hspace{-10pt}\braket{S_i^\mu} - \hspace{-10pt}\sum_{i\in \textrm{ blue sites}}\hspace{-10pt}\braket{S_i^\mu}\right)
\label{eqn:stagmag}
\eeq
where red and blue sites are those marked in Fig. \ref{fig:af_order}. Due to the different stripe orientations,
this gives us three stripe and zig-zag magnetizations for each spin-component $\mu$. In our notation $M_{z\textrm{-stripe}}^\mu$ corresponds to $M_{\textrm{staggered}}^\mu$ implemented for the stripe pattern shown in part (b) of Fig. \ref{fig:af_order}, as it aligns spins along $z$-bonds. Rotations of this pattern give $M_{x\textrm{-stripe}}^\mu$ and $M_{y\textrm{-stripe}}^\mu$.

\section{Calculation of multi-spin observables and correlators in HMFT}
\label{app:observables}
Our HMFT simulation has a wavefunction $\ket\Phi$ given by Eq. \eqref{eqn:cga} that can be expressed as a tensor product of identical single-cluster wavefunctions $\ket{\psi_\mathbf{R}}$.
As such, the expectation value of a multi-spin product with sites located within different clusters decomposes into the product of the expectation value within each cluster,
\begin{align}
\label{eqn:expt_val} \hspace*{-0.4cm}
\braket{\Psi|S_{\mathbf R, i}^\alpha S_{\mathbf{R},j}^\beta S_{{\mathbf R'},k}^\mu S_{{\mathbf R'}l}^\nu|\Psi} \nonumber \\
&\hspace*{-3.cm} =\braket{\psi_{\mathbf R}|S_{\mathbf R, i}^\alpha S_{\mathbf{R},j}^\beta |\psi_\mathbf{R}}
\braket{\psi_{\mathbf R'}|S_{{\mathbf R'},k}^\mu S_{{\mathbf R'}l}^\nu|
\psi_\mathbf{R'}}. 
\end{align}
This decomposition is relevant for two observables we calculate: plaquette flux $W_p$ and scalar chirality $\chi_{ijk}$. With our 6- and 24-site clusters, these operators can occur as single-cluster, two-cluster, and three-cluster terms. To average plaquette flux and chirality, we therefore need to appropriately decompose the observable as in Eq. \eqref{eqn:expt_val} and then sum them with appropriate weights ($1/2$ for two-cluster and $1/3$ for three-cluster terms) to avoid double counting terms belonging to more than one cluster. Specifically for chirality, the weights are applied as
\begin{eqnarray}
\overline \chi &=&
\frac{1}{N_c}\hspace{-5pt}\sum_{\substack{\langle\langle i,j,k\rangle\rangle\\\{i,j,k\}\in \mathbf R}} \hspace{-5pt}\braket{\chi_{ijk}} 
+ \frac{1}{2N_c}\hspace{-10pt}\sum_{\substack{\langle\langle i,j,k\rangle\rangle\\ \{i,j\} \in \mathbf R, k\in \mathbf R'}} \hspace{-10pt}\braket{\chi_{ijk}}\nonumber\\
&&+ \frac{1}{3N_c}\hspace{-15pt}\sum_{\substack{\langle\langle i,j,k\rangle\rangle\\ i\in \mathbf R, j \in \mathbf R', k \in \mathbf R''}}\hspace{-15pt}\braket{\chi_{ijk}} ,\label{eqn:chi_avg}
\end{eqnarray}
where 
$\langle\langle i,j,k\rangle\rangle$ refers to next-nearest neighbor sites of the honeycomb lattice forming triangles, and the first, second, and third sums corresponds to triangles with all three sites within a single cluster (at $\mathbf R$), those with sites split between the cluster $\mathbf R$ and neighboring clusters $\mathbf R'$, and those shared between three clusters ($\mathbf R$, $\mathbf R'$, and $\mathbf R''$), respectively.

Results computed using this method are shown in Figs. \ref{fig:summary} in the main text and Fig. \ref{fig:h0_results} in the preceding appendix.

\section{Calculation of topological properties in ED and HMFT}
\label{app:topo}

HMFT has not previously been applied to models with topological order, but our results in the KSL phase indicate the method correctly captures key topological properties of the model. As shown in Fig. (3) in the main text, at $h$=0 we find topological entanglement entropy consistent with exact results ($S_{\textrm{topo}}=-\log(2)$).

In addition to topological entanglement entropy, it is possible to calculate the many-body Chern number \cite{ortiz1994,fukui2005,varney2011}
and the topological $S$-matrix \cite{zhang2012} from ED simulations. We are unable, however, to perform those calculations in HMFT. Techniques for computing many-body Chern number require specific (twisted) boundary conditions incompatible with the mean-fields used in HMFT. To find the $S$-matrix, linear combinations of degenerate or quasi-degenerate wavefunctions belonging to the ground state manifold must be used, while HMFT provides access to only a single ground state. As such, we exclusively use ED to calculate these topological properties.

Details of how we obtained topological properties are illustrated in the following subsections.

\subsection{Topological entanglement entropy}

To find topological entanglement entropy $S_\text{topo}$, we take inspiration from previous work \cite{ronquillo2019} in using the Kitaev-Preskill (KP) construction \cite{kitaev2006a}:
First, we partition our system into four mutually connected subsystems $A$, $B$, $C$, and $D$.
Then, the topological entanglement entropy is given by
\beq
S_{\textrm{topo}} = S_A + S_B + S_C - S_{AB} - S_{BC} - S_{AC} + S_{ABC} ,
\eeq
where $S_A$ is the entanglement entropy acquired by tracing out degrees of freedom outside of region 
$A$, $S_{AB}$ is the entanglement entropy acquired by tracing out degrees of freedom outside of $A\cup B$, and so on.

\begin{figure}
\includegraphics[width=.5\linewidth]{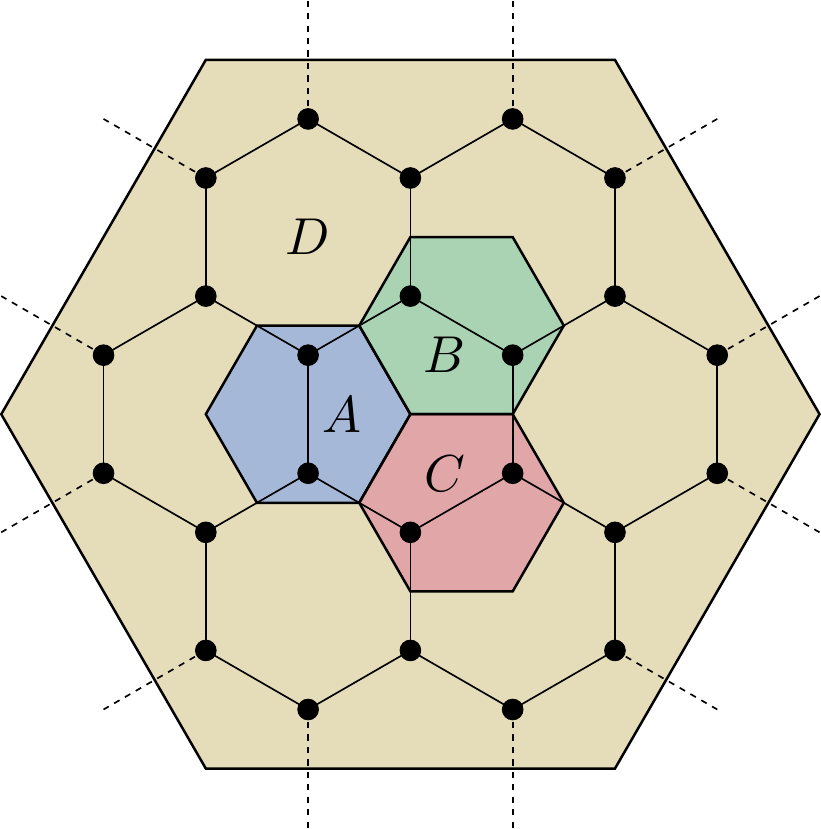}
\caption{Partitions used in the Kitaev-Preskill construction of the topological entanglement entropy from HMFT results. 
}
\label{fig:partitions}
\end{figure}

As this calculation occurs in the bulk of a cluster and has no reliance on boundary conditions, it can easily be performed in HMFT using the same techniques as ED, albeit with a restricted choice of partitions as compared to what is available when periodic boundary conditions are utilized. Since HMFT breaks quantum correlations at the cluster boundaries, partitions must be chosen to connect entirely within the bulk of the cluster, as shown in Fig. \ref{fig:partitions}. ED with PBC allows for larger partitions to be chosen \cite{ronquillo2019}.

The difference choice explains why ED finds an increase in the magnitude of $S_\textrm{topo}$ in the intermediate phase \cite{ronquillo2019}, while HMFT sees only a local increase compared to immediately adjacent regions, with a maximal value much lower in magnitude than the $-\ln 2$ recovered at $h$=0. The ED calculations are performed with larger partitions (illustrated in the supplemental material of \cite{ronquillo2019}). One possibility is that the smaller partitions required by the HMFT calculation are not sufficient to accommodate an increased correlation length in the intermediate phase.

\subsection{Topological \texorpdfstring{$\boldsymbol{S}$}{S}-matrix}

The topological $S$-matrix may be calculated in ED systems using an approach inspired by the KP topological entropy. By choosing partitions that bifurcate the cluster into disconnected regions and then finding linear combinations of the quasi-degenerate ground states to extremize entanglement entropies along these partitions, the topological $S$-matrix can be calculated via taking overlaps of these states \cite{zhang2012}.
These $S$-matrices may also be calculated from properties of idealized topological systems, such as the categories in Kitaev's 16-fold way \cite{kitaev2006}. As such, the $S$-matrix provides a useful quantity to to check for correspondence between ED on finite systems and topological quantum field theory results.

This approach has already been applied to the KSL in a magnetic field, finding results in good agreement with the exact ($h$=0) result \cite{hickey2019}. We are able to reproduce this result over a range of magnetic fields in the KSL region using PBC on the 24H cluster. This calculation was not attempted on the 24P or 18P cluster because these clusters recover an incorrect two-fold quasi-degeneracy in the KSL phase rather than the required three-fold quasi-degeneracy.

Outside of the KSL phase, the same calculation may be performed, but with an important caveat that the ED spectrum (see Fig. \ref{fig:ed_energy}) does not appear to exhibit the requisite topological (quasi)-degeneracy in any non-KSL phase. Instead, within each phase, level crossings occur between the low-lying excited states, meaning they do not represent a topologically protected manifold.
As such, the calculation is not well-motivated and the manifold of states used is arbitrary. Unsurprisingly, the $S$-matrices thus calculated do not conform to known topological quantum field theories and change drastically within the phases at points where low-lying excited states undergo level crossings. Along with the lack of topological degeneracy, this reinforces our understanding that whatever the nature of the intermediate phase is, it is not the gapped topological system suggested in \cite{zhang2021}, which should have a well-defined $S$-matrix given by Kitaev's 16-fold way.

\subsection{Many-body Chern number}

To calculate the many-body Chern number, twisted boundary conditions (TBC) are implemented on the 
6H and 24H cluster. For spin degrees of freedom, these boundary conditions are defined as
\beq \hspace*{-0.3cm}
S_{\mathbf{r}+\mathbf L_i}^+ = e^{i\phi}S_\mathbf{r}^+,
\quad
S_{\mathbf{r}+\mathbf L_i}^- = e^{-i\phi}S_\mathbf{r}^-,
\quad
S_{\mathbf{r}+\mathbf L_i}^z = S_\mathbf{r}^z ,
\eeq
where $\mathbf L_1$, $\mathbf L_2$ are vectors wrapping around the torus in PBC and $\phi_1$, $\phi_2$ are phases chosen while constructing the boundary conditions. When $\phi_1=\phi_2=0$, the
TBC reduce to PBC.

With boundary conditions established, we calculated the Chern number using a \emph{numerically gauge invariant} \cite{ortiz1994} formulation designed to cancel out any arbitrary phases present in the wavefunctions \cite{fukui2005,varney2011}: on an $L$$\times$$L$ grid of discrete phases 
$\phi_{1,2} \in \{0, \frac{2\pi}{L}, \dots, \frac{2\pi(L-1)}{L}\}$
with $\vec \phi = \{\phi_1, \phi_2\}$,
\beq
\widetilde C = \frac{1}{2\pi i}\sum_{\vec{\phi}} \ln \frac{U_1\l(\vec\phi\r) U_2\l(\vec \phi+\hat 1\r)}{U_1\l(\vec\phi+\hat 2\r) U_2\l(\vec \phi\r)},
\eeq
where $\hat 1 = \frac{2\pi}{L}\l(1,0\r)$,
$\hat 2 = \frac{2\pi}{L}\l(0,1\r)$. The variables
\beq
U_\mu = \frac{\braket{\vec \phi|\mathbf \phi+\hat \mu}}{|\braket{\vec \phi|\mathbf \phi+\hat \mu}|} 
\eeq
are defined at each point on the grid with
$\ket{\vec{\phi}}$ indicating the ground state of the Hamiltonian with TBC defined by $\vec \phi$.
Even for very coarse grids (small $L$), this formulation returns well-quantized integers,
and for large enough $L$, $\widetilde C$ corresponds to the continuum Chern number $C$ \cite{varney2011}.

Using this formulation on 6$\times$6 and 12$\times$12 grids, we find $C$=1 within the KSL phase, consistent with exact results \cite{kitaev2006}. In the PP phase, we find the unsurprising result that $C=0$, consistent with its trivial Landau order. As suggested by the lack of a gap closing between PP and $\chi$-PP phases in ED, we also obtain $C$=0 in the $\chi$-PP phase.\footnote{While vector chiral order is usually accompanied by a nonzero Chern number, the $\chi$-PP phase is characterized by \emph{scalar} chirality.}

In the intermediate phase, Chern number varies with $h$, jumping between different integers (quantized to our working numerical precision) and changing drastically with changing grid sizes. On the 18P cluster (whose smaller Hilbert space allows for much quicker calculations), even very fine grids (20$\times$20) did not resolve these non-physical changes in Chern number.
This phenomena could be an indication of gaplessness, which would prevent measurement of the Chern number in the thermodynamic limit. Further work is needed to clarify the meaning of these results.

\section{Brillouin zone of the honeycomb lattice}
\label{app:kspace}

\begin{figure}
\includegraphics[width=.8\linewidth]{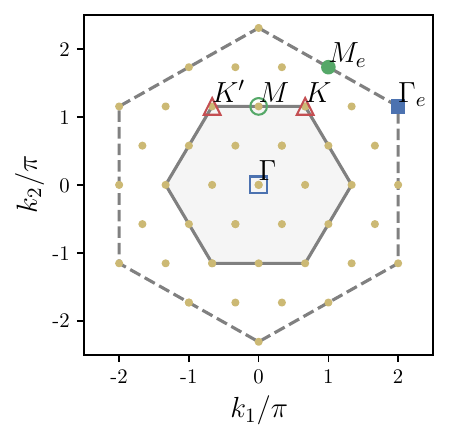}
\caption{Extended Brillouin zone of the honeycomb lattice showing points commensurate with the 24H cluster.
$\Gamma$, $K$, $K'$, and $\Gamma_e$ points are also commensurate with the 6-site cluster, while the other points (crucially, $M$ and $M_e$) are not.
Labels correspond to the same points utilized in Fig. \ref{fig:k_phase}.
}
\label{fig:bz}
\end{figure}

The honeycomb lattice is not a Bravais lattice. Rather, it consists of a triangular (Bravais) lattice of two-site unit cells. Because of this, some properties of the reciprocal lattice can be counterintuitive.

We can construct the honeycomb lattice with unit cell translation vectors
\begin{equation}
\mathbf a_1 = \frac{a}{2}\begin{pmatrix}1 \\ \sqrt{3}\end{pmatrix},
\quad
\mathbf a_2 = \frac{a}{2}\begin{pmatrix}-1 \\ \sqrt{3}\end{pmatrix},
\end{equation}
which have corresponding reciprocal lattice vectors
\begin{equation}
\mathbf b_1 =  2\pi \begin{pmatrix}1 \\ 1/\sqrt{3}\end{pmatrix},
\quad
\mathbf b_2 =  2\pi \begin{pmatrix} -1 \\ 1/\sqrt{3}\end{pmatrix}.
\end{equation}
The vectors $\mathbf a_i$ connect next-nearest neighbors (sites belonging to the same triangular sublattice).
To complete the honeycomb lattice, we require a third vector to connect sites of opposite sublattices. One choice is
\beq
\mathbf a_3 = \frac{1}{\sqrt 3}\begin{pmatrix} 0 \\ 1 \end{pmatrix}.
\eeq

\begin{figure}
\includegraphics[width=1.\linewidth]{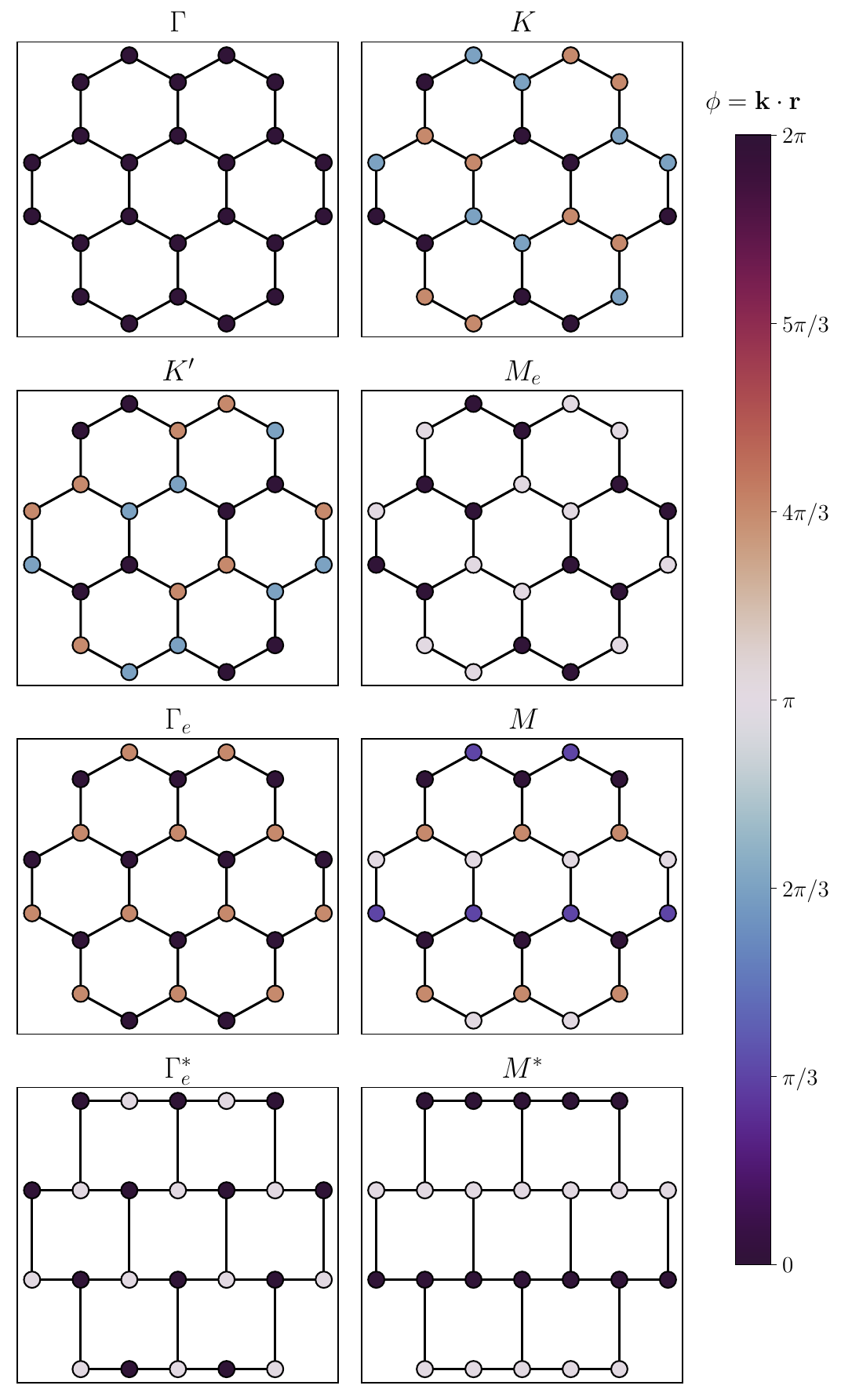}
\caption{Phase $\phi=\mathbf r \cdot \mathbf k$ at each point in the 24-site cluster
for various $\mathbf k$ corresponding to different local orders.
The $M$ and $\Gamma_e$ plots are plotted on both honeycomb and brickwall lattices ($M^*$, $\Gamma_e^*$). Note that
only on the brickwall lattice do these wavevectors produce the $\pi$ relative phases corresponding to zig-zag and N\'eel order.}
\label{fig:k_phase}
\end{figure}

Because of this structure within unit cells, points in $\mathbf k$-space outside of the first Brillouin zone (BZ) of the underlying triangular lattice correspond to different phases $\mathbf r \cdot \mathbf k$, and therefore have different physical meaning. As such, we construct the \emph{extended Brillouin zone} to accommodate these additional points. This extended BZ is depicted in Fig. \ref{fig:bz}, with high-symmetry points labelled (and with a subscript $e$ indicating points outside the first BZ).

Figure \ref{fig:k_phase} shows the phase $\mathbf r \cdot \mathbf k$ acquired at each point in the 24H cluster at high-symmetry wavevector. From this, it is clear that the $\Gamma$ wavevector corresponds to a uniform magnetization (as expected) and $M_e$ forms stripe order (along $y$-bonds in this case, with the different $M_e$ points resulting in different stripe orientations). 

On the other hand, the $\Gamma_e$ point on the honeycomb lattice has nearest-neighbor spins acquiring a relative phase of $2\pi/3$. If this phase were instead $\pi$, this would correspond to N\'eel order. Surprisingly, no single wavevector on the honeycomb lattice assigns a relative $\pi$ phase to nearest neighbors. Instead, to find wavevector corresponding to N\'eel order, we deform the honeycomb lattice into the topologically equivalent brickwall lattice (as in the lowest two subplots in Fig. \ref{fig:k_phase}). Under such a deformation, the $\Gamma_e$ point does correspond to N\'eel order. Similarly, the $M$ points go from producing a variety of relative phases on the honeycomb lattice to producing a $\pi$ relative phase on sites corresponding to zig-zag order on the brickwall lattice. For this reason, the structure factors $S(M)$ and $S(\Gamma_e)$ plotted in Fig. \ref{fig:ed_stripe}(b) were computed on the brickwall lattice.

\end{document}